\documentclass[aps,prd,twocolumn,superscriptaddress,floatfix,letterpaper,nofootinbib]{revtex4-1}

\usepackage{graphicx}

\usepackage[usenames, dvipsnames]{color}
\usepackage[colorlinks,citecolor=blue,linkcolor=blue,urlcolor=blue]{hyperref}

\usepackage{amsmath, amsthm, amssymb,slashed}

\pdfoutput=1

\usepackage[normalem]{ulem}

\usepackage{booktabs}

\definecolor{mygray}{gray}{0.6}

\usepackage{upgreek}
\usepackage{bbm}

\newenvironment{myfont}[2][]{\csname#2\endcsname[#1]}{}

\usepackage{slashed}
\usepackage[makeroom]{cancel}
\usepackage[normalem]{ulem}
\usepackage{soul}
\newcommand{\stkout}[1]{\ifmmode\text{\sout{\ensuremath{#1}}}\else\sout{#1}\fi}

\usepackage{sseq}
\usepackage[all,cmtip]{xy}
\usepackage{tikz-cd}
\usepackage{tikz}
\usetikzlibrary{matrix}
\usetikzlibrary{decorations.markings}

\newcommand{\bea}{\begin{eqnarray}}
\newcommand{\eea}{\end{eqnarray}}
\def\be{\begin{equation}}
\def\ee{\end{equation}}

\newcommand{\e}{\hspace{1pt}\mathrm{e}}

\newcommand{\ii}{\hspace{1pt}\mathrm{i}\hspace{1pt}}

\def\CP{{\mathbb{CP}}}

\definecolor{red}{rgb}{1,0,0}
\definecolor{blue}{rgb}{0,0,1}
\definecolor{dblue}{rgb}{0,0,0.4}
\definecolor{green}{rgb}{0,1,0}
\definecolor{black}{rgb}{0,0,0}
\definecolor{white}{rgb}{1,1,1}

\definecolor{brn}{rgb}{.8,.4,.0}
\definecolor{redo}{rgb}{1,.5,.0}
\definecolor{ddgrn}{rgb}{0,0.4,0}
\definecolor{dgrn}{rgb}{0,0.55,0}
\definecolor{dbl}{rgb}{0,0,0.5}

\usepackage[bbgreekl]{mathbbol}
\usepackage{amscd}

\newcommand{\Z}{\mathbb{Z}}
\newcommand{\C}{\mathbb{C}}

\newcommand{\Ref}[1]{Ref.~[\onlinecite{#1}]}

\newcommand{\eq}[1]{eq.~(\ref{#1})}

\newcommand{\Tr}{{\rm Tr}} 
 
\renewcommand{\Im}{{\rm Im}} 
\renewcommand{\Re}{{\rm Re}}

\newcommand{\bpm}{\begin{pmatrix}}
\newcommand{\epm}{\end{pmatrix}}
\newcommand{\bmm}{\begin{matrix}}
\newcommand{\emm}{\end{matrix}}

\newcommand{\al}{\alpha} 
\newcommand{\bt}{\beta}

\def\CD{{\cal D}}

\def\CM{{\cal M}}
\def\CN{{\cal N}}

\def\Z{{\mathbb{Z}}}

\def\C{{\mathbb{C}}}

\def\Tr{{\mathrm{Tr}}}

\def \Hom{\operatorname{Hom}}

\def \Im{\operatorname{Im}}
\def \H{\operatorname{H}}

\def \Z{\mathbb{Z}}
\def \A{\mathcal{A}}

\def \CP{\mathbb{CP}}

\def\Ext{\operatorname{Ext}}

\newcommand\hcup[1]{\underset{{\scriptscriptstyle #1}}{\cup}}

\newcommand {\emptycomment}[1]{}

\def\Sq{\mathrm{Sq}}
\def\B{\mathrm{B}}

\newcommand{\SO}{{\rm SO}}
\newcommand{\Spin}{{\rm Spin}}
\newcommand{\U}{{\rm U}}
\newcommand{\SU}{{\rm SU}}

\newcommand{\Pin}{{\rm Pin}}
\newcommand{\tO}{{\rm O}}

\newcommand{\tA}{{\rm A}}

\newcommand{\nn}{{\nonumber}}

\usepackage{datetime}

\newcommand{\Fig}[1]{Fig.~\ref{#1}}

\newcommand{\Sec}[1]{Sec.~\ref{#1}} 
\newcommand{\App}[1]{Appendix \ref{#1}} 
\usepackage{enumitem,xcolor,cleveref}

\hypersetup{
  colorlinks = true,
  linkcolor = blue
}

\begin{document}

\begin{titlepage}


\title{Adjoint QCD$_4$, Deconfined Critical Phenomena,\\[2mm] 
Symmetry-Enriched 
Topological Quantum Field Theory,\\[2mm] 
and Higher Symmetry-Extension}

\author{Zheyan Wan}
\affiliation{School of Mathematical Sciences, USTC, Hefei 230026, China}

\author{Juven Wang}
\email{juven@ias.edu}
\affiliation{School of Natural Sciences, Institute for Advanced Study, Einstein Drive, Princeton, NJ 08540, USA }
\affiliation{Center of Mathematical Sciences and Applications, Harvard University, MA 02138, USA}


\begin{abstract}

Recent work explores the candidate phases of the 4d adjoint quantum chromodynamics (QCD$_4$) with an SU(2) gauge group and two massless adjoint Weyl fermions.
Both Cordova-Dumitrescu and Bi-Senthil 
propose possible low energy 4d topological quantum field theories (TQFTs) 
to saturate the higher 't Hooft anomalies of adjoint QCD$_4$ under a renormalization-group (RG) flow from high energy.
In this work, we generalize the symmetry-extension method of Wang-Wen-Witten [arXiv:1705.06728, Phys.~Rev.~X 8, 031048 (2018)] 
to higher symmetries, and formulate  
higher group cohomology and cobordism theory approach 
to construct higher-symmetric TQFTs.
We prove that the symmetry-extension method saturates certain anomalies, but also prove that \emph{neither} 
$A \mathcal{P}_2(B_2)$ \emph{nor}  $\mathcal{P}_2(B_2)$ can be fully trivialized,
with the background 1-form field $A$, Pontryagin square $\mathcal{P}_2$ and 2-form field $B_2$.  
Surprisingly, 
this indicates an obstruction to constructing a
fully 1-form center and 0-form chiral symmetry (full discrete axial symmetry) preserving 
4d TQFT with confinement, 
a \emph{no-go} scenario via symmetry-extension for specific higher anomalies.
We comment on the implications and constraints on \emph{deconfined quantum critical points} (dQCP), 
\emph{quantum spin liquids} (QSL) or \emph{quantum fermionic liquids}  in condensed matter, {and ultraviolet-infrared (UV-IR) duality} in 3+1 spacetime dimensions.


\end{abstract}

\pacs{}

\maketitle

\end{titlepage}

\tableofcontents

\section{Introduction and Summary of Main Results}
\label{sec:intro}

Recent work explores the candidate phases of the adjoint quantum chromodynamics in 4 dimensional spacetime (QCD$_4$) 
with an SU(2) gauge group and two massless adjoint Weyl fermions
(equivalently, two massless adjoint Majorana fermions, or one massless adjoint Dirac fermion)
\cite{Anber2018tcj1805.12290,2018arXiv180609592C,Bi:2018xvr, SWW}.\footnote{
In this case, we denote the adjoint Weyl fermion flavor $N_f =2$ and the gauge group $N_c=2$ for SU($N_c$).}
This adjoint QCD$_4$ has a 1-form electric $\Z_2$ center global symmetry,
which is a generalized global symmetry of higher differential form \cite{Gaiotto:2014kfa}.
This adjoint QCD$_4$ has the SU(2) gauge theory coupling to the matter fields in the adjoint representation, thus
it gains a 1-form electric $\Z_2$ center symmetry;
while the usual fundamental QCD$_4$ has the gauge theory coupling to the matter fields in the fundamental representation, which
lacks the 1-form symmetry.
We will soon learn that this 1-form symmetry plays a crucial rule to constrain the higher 't Hooft anomaly-matching \cite{tHooft:1980xss}
of the quantum phases of the adjoint QCD$_4$.
(See \Sec{sec:thy} for more detailed information regarding the global symmetries and 't Hooft anomalies of this adjoint QCD$_4$.)

Given the adjoint QCD$_4$ at the high energy scale, 
it is known that this theory is weakly coupled 
thus asymptotically free at ultraviolet (UV free) when the number of Weyl fermion flavor $N_f \leq 5$.
Viewing the adjoint QCD$_4$ as a UV completion of a quantum field theory (QFT), 
we should ask what this QFT flows to
under a renormalization-group (RG) flow from ultraviolet (UV)
to the low energy at infrared (IR).
Both  
Cordova-Dumitrescu \cite{2018arXiv180609592C} 
and
Bi-Senthil \cite{Bi:2018xvr} propose its low energy 
candidate phases at IR,
saturating the higher 't Hooft anomalies involving the 1-form symmetry.

In particular,
Bi-Senthil \cite{Bi:2018xvr} suggests a fully-symmetric 
4d TQFT 
to saturate higher 't Hooft anomalies without breaking any
UV global symmetries of the adjoint QCD$_4$.
Namely, an interesting RG flow from Bi-Senthil \cite{Bi:2018xvr} speculates that:
\bea \label{eq:dual}
&&\text{{Adjoint QCD$_4$ at UV }} {\overset{\text{RG flow (?) to a long distance}}{{ \xrightarrow{\hspace*{3.75cm}}}}} \\
&&\text{Massless 1 Dirac (2 Weyl) fermion + 4d TQFT at IR (?)}.\nn
\eea\\

The IR theory only involves a massless free 1 Dirac (or 2 Weyl) fermion,
and a decoupled 4d TQFT. Since the massless 1 Dirac fermion has 
only the ordinary 0-form symmetry 
but \emph{no} 1-form symmetry, so the 
massless fermion sector alone \emph{cannot} saturate the higher anomaly of the adjoint QCD$_4$.
Thus the crucial and nontrivial check on Bi-Senthil \cite{Bi:2018xvr} proposal of this UV-IR duality \eq{eq:dual} relies
on the explicit construction of the fully-symmetric 4d TQFT to saturate all
higher 't Hooft anomalies involving 1-form symmetry.
One of the motivation of our present work is to
rigorously verify the validity of this symmetric anomalous 4d TQFT.

In this work, we have two goals:
\begin{enumerate}

\item
We generalize the symmetry-extension method of Wang-Wen-Witten  \cite{Wang2017loc170506728} to higher symmetries. 
We formulate a \emph{higher group cohomology} or a \emph{higher cobordism theory} approach of \Ref{Wang2017loc170506728}  
to construct ``symmetric anomalous TQFTs'' that can live on the boundary of symmetry protected topological states (SPTs).
The ``symmetric anomalous TQFTs'' is an abbreviation of 
``the TQFTs that saturates the (higher) 't Hooft anomalies of a given global symmetry
by preserving the global symmetry.''
Previous works in condensed matter physics suggest that the
\emph{long-range entangled} anomalous topological order (whose effective low energy theory is a TQFT)
can live on the boundary of a \emph{short-range entangled} SPT state, see \cite{Senthil1405.4015} and References therein on this exotic phenomenon.
The boundary of SPTs protected by symmetry group $G$ (called $G$-SPTs) has the 't Hooft anomaly of symmetry $G$.
Ref.~\cite{Wang2017loc170506728} 
provides a systematic way to construct the
symmetric anomalous TQFTs for a $G$-SPTs of a given symmetry $G$.
In particular, among other results,
Ref.~\cite{Wang2017loc170506728} proves that:

``For any bosonic $G$-SPTs protected by a finite group $G$ 
(unitary or anti-unitary time-reversal symmetry) in a 2-dimensional spacetime  (2d) or above ($\geq 2d$), there \emph{always exists} a finite group $K$ bosonic gauge theory which is a 
TQFT, saturating the $G$-'t Hooft anomaly, 
that can live on the boundary of  $G$-SPTs, based on the symmetry-extension
method via a short exact sequence $1 \to K \to H\to G\to 1$,
where all $G,K$ and $H$ are finite groups of 0-form symmetry.''

In this article, we will explore the related phenomenon of Ref.~\cite{Wang2017loc170506728} but we improve 
the formulation by replacing the 
0-form $G$ symmetry to include generalized higher symmetries of Ref.~\cite{Gaiotto:2014kfa}.

\item
We apply the above generalized higher symmetry-extension method from Ref.~\cite{Wang2017loc170506728}
 either to construct the higher-symmetric anomalous TQFTs, for adjoint QCD$_4$; or to show the invalidity of the TQFTs via a symmetry-extension method.

Specifically, 
we find an \emph{obstruction} to construct certain symmetric 4d TQFTs via symmetry extension, for
the mixed anomaly mixing between the discrete axial symmetry (here the 0-form $\Z_{2 N_c N_f}$ = $\Z_{8}$ symmetry, with $N_c = N_f =2$) 
and the 1-form electric center symmetry (denoted as $\Z_{2,[1]}^e = \Z_{2,[1]}$).
This higher anomaly is abbreviated as the Type I higher anomaly in Ref.~\cite{Bi:2018xvr}.
The {Type I anomaly} in 4d has a $\Z_4$ class (below {$k \in \Z_4$ class}),
one can explicitly write down the 5d topological (abbreviated as ``topo.'') invariant \cite{2018arXiv180609592C} which is a cobordism invariant 
(see mathematical details in \cite{W2} and \Sec{sec:cobordism-TQFT}),
\be \label{eq:type1}
\text{Type I anomaly/topological invariant}:\quad \e^{\ii \frac{k\pi}{2} \int A \cup \mathcal{P}_2(B_2)}.\quad
\ee
{Here $A$ is the $\Z_4$-valued background 1-form gauge field coupling to the 0-form $\Z_8/\Z_2^F=\Z_4$ part of the axial global symmetry.
The $\Z_2^F$ is the fermionic parity symmetry which is $(-1)^{N_F}$, assigning a minus to the state of system when there is an odd number of total number of fermions ${N_F}$.
The $B_2$ is the $\Z_2$-valued background 2-form gauge field  coupling to the 1-form $\Z_{2,[1]}^e$-symmetry.
The $\cup$ is the cup product, and the $\mathcal{P}_2$ is the Pontryagin square, see more details in \Sec{sec:thy}.}
In \Sec{sec:cobordism-TQFT}, we will prove the non-existence of anomalous symmetric 4d TQFTs (of finite groups or higher groups) 
for this 4d higher anomaly (or equivalently, 5d higher SPTs) of \eq{eq:type1},
via the symmetry extension method. 
However, we clarify that our proof does not necessarily imply a no go theorem for the anomalous symmetric 4d TQFTs for Bi-Senthil \cite{Bi:2018xvr} in general, 
it could be due to the limitation of the symmetry extension \cite{Wang2017loc170506728} we used.
Nevertheless, it is known that \cite{Wang2017loc170506728}'s method is general and systematic enough to 
construct symmetric TQFT for all bosonic anomalies of the ordinary 0-form finite group symmetries;
thus the obstruction from \cite{Wang2017loc170506728} is severe and interesting by itself to be presented here. 
This proof indicates a no-go scenario for anomalous symmetric 4d TQFTs if we \emph{only} limit the construction 
under the symmetry-extension construction of TQFTs.

In contrast,
we find that the generalized symmetry-extension method 
can indeed construct another symmetric 4d TQFT saturating 
a different higher mixed anomaly,
mixing between the background gravity (or the curved spacetime geometry) 
and the 1-form center symmetry (denoted as $\Z_{2,[1]}$).
This higher anomaly is abbreviated as the Type II higher anomaly in Ref.~\cite{Bi:2018xvr}.
We can explicitly write down the 5d topological (abbreviated as ``topo.'') invariant \cite{2018arXiv180609592C} as the following cobordism invariant (see mathematical details in \cite{W2} and \Sec{sec:cobordism-TQFT}),
\begin{multline}  \label{eq:type2}
\text{Type II anomaly/topological invariant}:\\ 
 \e^{\ii {\pi}  \int w_2(TM) \Sq^1 B_2}=\e^{\ii {\pi}  \int w_3(TM) B_2}.
\end{multline}
Here  $w_j(TM)$ has the $w_j$ as the $j$-th Stiefel-Whitney (SW) class  \cite{milnor1974characteristic},
as the probed background spacetime $M$ connection over the spacetime tangent bundle $TM$. 
The $ \Sq^1$ is the Steenrod operation.
We demonstrate the explicit construction of the 4d symmetric anomalous 
TQFT for this 4d higher anomaly (or equivalently, 5d higher SPTs) of \eq{eq:type2} 
in \Sec{sec:higher-sym-ext}.

\end{enumerate}

Physically, the above description concerns the physics side of the story, relating to quantum field theory, QCD or the strongly-correlated systems in condensed matter physics.

Mathematically, we ask the following questions (corresponding to the physics story above)
and find an obstruction to a positive answer for a Bi-Senthil's scenario \cite{Bi:2018xvr} via the symmetry-extension alone, 
 generalizing the method of  \cite{Wang2017loc170506728}:
\begin{enumerate}[label=\textcolor{blue}{Question \arabic*.}, ref={Question \arabic*},leftmargin=*]
\item \label{que1}
Can we trivialize the topological term $A \cup\mathcal{P}_2(B_2)$ via extending the global symmetry by 0-form symmetry and 1-form symmetry?
To answer this, 
we deal with the trivialization problem of the cobordism invariant $A \cup\mathcal{P}_2(B_2)$ 
of the bordism group $\Omega_5^{\Spin\times_{\Z_2}\Z_8}(\B ^2\Z_2)$.\footnote{
In this work, we will use the term $d$d ``cobordism invariant'' to describe the $d$d topological term or $d$d (higher) SPTs.
On a manifold with boundary, the boundary of such a cobordism invariant (or SPTs) has a 't Hooft anomaly.
We denote the bordism group $\Omega_d^G$, while we denote the cobordism group $\Omega^d_G$.}
We prove that the answer is negative. 
\item \label{que2}
We also solve the trivialization problem of the cobordism invariant $\mathcal{P}_2(B_2)$ of the bordism group $\Omega_4^{\SO}(\B ^2\Z_2)$:
Can we trivialize the topological term $\mathcal{P}_2(B_2)$ via extending the global symmetry by 0-form symmetry and 1-form symmetry?
We prove that the answer is also negative. 
\end{enumerate}

The plan of the article goes as follows.

In \Sec{sec:thy}, we detail the related global symmetries and higher anomalies relevant for our goal,
following a remarkable Ref.~\cite{2018arXiv180609592C}.

In \Sec{sec:higher-sym-ext}, we discuss the higher {symmetry-extension} generalization of \cite{Wang2017loc170506728},
and successfully apply the method to construct a 4d symmetric anomalous TQFT for Type II anomaly \eq{eq:type2}.
But this method shows an obstruction for the Type I anomaly \eq{eq:type1}.

We leave rigorous but more formal and mathematical details of the calculation in Appendices.

In \App{sec:cobordism-TQFT}, we find a potential obstruction: 
The {Type I anomaly} \eq{eq:type1} cannot be saturate by a symmetric anomalous finite group/higher group TQFT,
at least by a {symmetry extension} method.

In \App{sec:proof}, we give a counter example as the proof for the failure of the {symmetry extension} method
applying to trivializing the 5d $A\cup\mathcal{P}_2(B_2)$.

In \App{sec:P(B)}, we show a similar obstruction: 
The 4d $\mathcal{P}_2(B_2)$ cannot be saturated by a symmetric anomalous finite group/higher group TQFT,
at least by a {symmetry extension} method. 

We note that the \App{sec:cobordism-TQFT}, \ref{sec:proof}, and \ref{sec:P(B)} are more technical and mathematical demanding.
For readers who are not familiar with the mathematical background for these three sections, one can either consult \cite{W2} and 
\cite{2017arXiv171111587GPW} (e.g. the Appendix of \cite{2017arXiv171111587GPW}),
or simply skip them and proceed to the conclusion \Sec{sec:conclude} which we summarize the physics interpretations of the above three
sections.

We conclude in \Sec{sec:conclude}. 

The mathematical details of our cobordism calculations can be found in a companion paper \cite{W2}.

\section{Theory of Adjoint QCD$_4$}
\label{sec:thy}

We have {an SU(2) gauge theory coupled to  2 $\mathbf{3}$ ($N_f=2$ for the 2, and the $\mathbf{3}$ for the triplet) adjoint Weyl fermions}
in the adjoint representation of SU(2).
The path integral (or partition function) of this adjoint QCD$_4$, in the Minkowski signature, viewed as a UV QFT theory can be written as:
\begin{widetext}
\onecolumngrid
\bea \label{eq:ZUV}
Z_{\text{UV}}&=&\int [\CD \psi] [\CD \bar \psi] [\CD a] \exp(\ii S_{\text{UV}}), \\
S_{\text{UV}}&=&\int d^4 x \;
\sum_{j=1,2}
\frac{\ii}{g^2} \bar \psi^{b'}_j  \bar \sigma^\mu (\partial_\mu - \ii g a_\mu^{a'} (T^{a'})_{b' b}) \psi^{j b} 
-  \frac{1}{g^2}\int  \text{Tr}(F \wedge \star F)  + \dots . \label{eq:SUV}
\eea
\end{widetext}
The \eq{eq:SUV} contains the first term as the Dirac Lagrangian, and the second term as the Yang-Mills Lagrangian. 
The $[\CD \dots]$ is the path integral measure for the quantum fields.
The $\bar \sigma^\mu \equiv (1, - \vec \sigma)$ contains the standard Pauli sigma matrices $\vec \sigma$.
Here the Weyl fermion $\psi^{j b}_\alpha$ has:\\
$\bullet$ the flavor index $j$ (of the classical U(2) flavor symmetry, or more precisely the $\frac{\SU(2) \times \Z_8}{\Z_2^F}$ flavor symmetry in a quantum theory, see later),\\
$\bullet$ the gauge index $b$ of the gauge SU(2) of adjoint triplet,\\
$\bullet$ the Lorentz index $\alpha$ of the Lorentz group.\\
The hermitian conjugation of fermion field is $\bar \psi^{b'}_j$ $=\psi^{j b' \dagger}$.
With the Lorentz index, we have $\bar \psi^{b'}_{\dot{\alpha} j}$ $=\psi^{j b' \dagger}_{\alpha}$, following the standard supersymmetry notation.\\
Here are some other comments:\\
$\bullet$ The $g$ is the dimensionless Yang-Mills coupling, which is a running coupling in the quantum theory.\\
$\bullet$ The $F$ is the SU(2) gauge field $a$'s 2-form field strength. The $\star F$ is the $F$'s Hodge dual. 
\\
$\bullet$ One can consider the deformation of the theory as extra terms in the $\dots$, 
such as the mass deformation \cite{SWW}, e.g. $({m_{ij}}{}  \delta_{b' b} \psi_\alpha^{i b'} \epsilon^{\al \bt}  \psi_\bt^{jb} + c.c.)$.
In the classical theory, we can add the $\theta$-term, 
\bea \label{eq:theta}
\int\limits_{} (\frac{  \theta}{8 \pi^2}  \text{Tr}\,F \wedge F).
\eea
However, in the quantum theory, with the presence of the fermion fields $\psi$, we can rotate the $\theta$ away. 
If we have the mass term for the fermions, we can absorb the $\theta$-term into the complex fermion mass matrix in the mass deformation.\\

\subsection{Global Symmetries}
The global symmetries of the adjoint QCD$_4$ \eq{eq:ZUV} has been analyzed systematically in 
\cite{2018arXiv180609592C}. Here we recap the results and will write the results suitable for the cobordism theory analysis later in Appendices
\ref{sec:cobordism-TQFT} to \ref{sec:P(B)}.
\begin{enumerate}

\item Flavor symmetry $\frac{\SU(2) \times \Z_8}{\Z_2^F}$: The classical flavor symmetry of 2 triplet Weyl fermions is the flavor $\U(2)=\frac{\SU(2) \times \U(1)_\tA}{\Z_2^F}$.
However, the axial symmetry $\U(1)_\tA$ is broken down to a discrete axial symmetry $\Z_{2 N_c N_f,\tA}$, which is $\Z_{2 N_c N_f,\tA}=\Z_8$ here,
due to the Adler-Bell-Jackiw (ABJ) anomaly.\footnote{For a clarification of different meanings of anomalies,
such as the three different types of physics of anomalies:
(1) Classical global symmetry is violated at the quantum theory: ABJ anomaly.
(2) Quantum global symmetry is well-defined and preserved but with the 't Hooft anomaly.
(3) Dynamical gauge anomaly;
the readers can consult, for example, the Section 1 Introduction of  \cite{Wan2018zql1812.11968}
and References therein.} 
It is a standard calculation of
 the $U(1)_\tA$-axial symmetry is explicitly broken by the dynamical $\SU(N_c)$-gauge instanton down to
$\Z_{2N_cN_f,\tA}$-axial symmetry.

So the flavor symmetry is simply $\frac{\SU(2) \times \Z_{8,\tA}}{\Z_2^F}= \frac{\SU(2) \times \Z_8}{\Z_2^F}$ for the quantum theory.
The $\SU(2)$ is also written as the $\SU(2)_R$ as the $R$-symmetry thanks to the standard convention in $\CN=2$ supersymmetric Yang-Mills theory (SYM) \cite{Seiberg1994rs9407087}.  
In the $\CN=2$ SYM, the adjoint fermions are gauginos.

\item The 1-form center symmetry $\Z_{2,[1]}^e \equiv \Z_{2,[1]}$: The adjoint QCD has the matter in adjoint representation, so the SU($N_c$) (here SU(2))
fundamental Wilson line is charged under the 1-form electric center symmetry $\Z_{2,[1]}^e$ measured by a 2-surface ``charge'' operator.
The ``charged'' fundamental Wilson line (spin-1/2 representation of SU(2)) has an odd $\Z_2$ charge.
The odd half integer spin-$n$/2 representation of SU(2) has an odd $\Z_2$ charge of 1-form symmetry.
Wilson lines of other integer spin-$n$ representation (e.g. the adjoint) of SU(2) has a trivial (namely even) $\Z_2$ charge of 1-form symmetry.

Importantly the 1-form center symmetry $\Z_{2,[1]}^e$ is preserved means that the electric Wilson loop ($e$-loop) is unbreakable,
or called tension-ful \cite{Bi:2018xvr}. 
Since the adjoint QCD has the 1-form center symmetry,
we can use the 1-form center symmetry charged object to detect: \\
$\bullet$ Confinement: If 1-form symmetry is preserved, and all the Wilson loops (of all representations) obey the area law.\\
$\bullet$ Deconfinement: If 1-form symmetry is spontaneously broken, then the 
 Wilson loops of odd half integer spin-$n$/2 representation (e.g. fundamental representation) obey the perimeter law.

\item Spacetime symmetry: In Lorentz signature, we have the Poincar\'e group symmetry which contains the Lorentz group.
We also have the discrete $CPT$ symmetries. There is no charge conjugation $C$ for SU(2) gauge theory due to
the lack of SU(2) outer automorphism. So there is only $T$ and $P$ symmetry interchangeably thanks to the $CPT$ theorem.
If we focus on orientable spacetime for the adjoint QCD in $d$d, we can consider the $\Spin(d)$ spacetime symmetry, for the purpose of classifying the 't Hooft anomalies through the
cobordism theory \cite{Freed2016, W2}.
If we consider the non-orientable spacetime for the adjoint QCD in $d$d, we should consider the $\Pin^-(d)$ spacetime symmetry, for the purpose of classifying the 't Hooft anomalies through a cobordism theory, See Ref.~\cite{Freed2016, W2}.
This adjoint QCD is a fermionic theory,
the spacetime symmetry ${G_{\text{spacetime} }}$ and the internal symmetry ${\mathbb{G}_{\text{internal}} }$ shares the fermionic parity $\Z_2^F$,
so the precise way to write the full global symmetry would be:
\bea
({\frac{{G_{\text{spacetime} }} \times  {\mathbb{G}_{\text{internal}} }}{\Z_2^F}})
\equiv
{{G_{\text{spacetime} }} \times_{\Z_2^F}  {\mathbb{G}_{\text{internal}} }},
\eea
where the common ${\Z_2^F}$ is mod out, while the ``$\times_{\Z_2^F}$'' notation follows \cite{Freed2016}.

\end{enumerate}

By combining the internal global symmetry (flavor and 1-form center symmetries) and the spacetime global symmetry above, 
the overall global symmetry can be written as:
\bea \label{eq:Spin-sym}
{\Spin \times_{{\Z_2^F}} \big(\frac{\SU(2) \times \Z_{8,\tA}}{\Z_2^F} \big) \times  \Z_{2,[1]}^e}
\eea
\bea \label{eq:Pin-sym}
{\Pin^- \times_{{\Z_2^F}} \big(\frac{\SU(2) \times \Z_{8,\tA}}{\Z_2^F} \big) \times  \Z_{2,[1]}^e}
\eea

Below we follow Ref.~\cite{W2}, which generalizes a theorem in a remarkable work of Freed-Hopkins \cite{Freed2016}.
Freed-Hopkins \cite{Freed2016} formulates a cobordism theory  --- whose cobordism group, of the ordinary 0-form global symmetries, classifies a class of symmetric invertible TQFTs,
which is relevant to the SPT classification.
Ref.~\cite{W2} generalizes \cite{Freed2016} to a cobordism theory of the higher global symmetries (e.g. including 0-form global symmetries and 1-form global symmetries)
and computes some examples of such cobordism groups.

In terms of bordism group notation, which later will be helpful for identifying all the (higher) 't Hooft anomalies
and the SPT classes via the computations of \cite{W2}, we write their corresponding bordism groups 
$\Omega_{d}$ as:\footnote{
Here $\B G$ means the classifying space of $G$, and $pt$ means the point.}
\begin{itemize}

\item  Bordism group for \eq{eq:Spin-sym}:
\begin{multline}
\label{eq:Spin-cob}
\Omega_{d}^{\Spin \times_{{\Z_2^F}} \big(\frac{\SU(2) \times \Z_{8,\tA}}{\Z_2^F} \big) \times \B  \Z_{2,[1]}^e}(pt)\\
\equiv 
\Omega_{d}^{\Spin \times_{{\Z_2^F}} \big(\frac{\SU(2) \times \Z_{8,\tA}}{\Z_2^F} \big) }
(\B^2  \Z_{2,[1]}^e).
\end{multline}

\item Bordism group for \eq{eq:Pin-sym}:
\begin{multline}
\label{eq:Pin-cob}
\Omega_{d}^{\Pin^- \times_{{\Z_2^F}} \big(\frac{\SU(2) \times \Z_{8,\tA}}{\Z_2^F} \big) \times \B  \Z_{2,[1]}^e}(pt)\\
\equiv 
\Omega_{d}^{\Pin^- \times_{{\Z_2^F}} \big(\frac{\SU(2) \times \Z_{8,\tA}}{\Z_2^F} \big) }
(\B^2  \Z_{2,[1]}^e).
\end{multline}
\end{itemize}
For adjoint QCD$_4$ in 4d, the higher 't Hooft anomalies are classified by
the dimension $d=5$ for the above 
bordism groups.\footnote{On the other hand, if we aim to know the 4d SPTs compatible with the symmetry of adjoint QCD$_4$,
then we need to consider the dimension $d=4$ for the above bordism groups.
This research direction is pursued by \Ref{2017arXiv171111587GPW} for the related
SU($N_c$) Yang-Mills gauge theories.}
See more details in  \Ref{Wan2018zql1812.11968}.

\subsection{Anomalies}
\label{sec:anom}

Now consider the $d=5$ bordism groups above in \eq{eq:Spin-cob} and \eq{eq:Pin-cob},
we like to match their selective 5d cobordism invariants to
 the anomalies captured by the 4d adjoint QCD$_4$.
 
Cordova-Dumitrescu  \cite{2018arXiv180609592C} have captured several anomalies, which we now overview:
\begin{enumerate}

\item The SU(2) Witten anomaly \cite{Witten:1982fp} 
for the flavor SU(2)$_R$ sector, due to that there is an odd number of SU(2)$_R$ flavor doublet.
The appearance of SU(2) Witten anomaly also indicates the IR fate of this adjoint QCD$_4$ is gapless instead of fully gapped.

\item The $(\Z_{8,\tA})^3$ anomaly captured by a perturbative anomaly (i.e., a triangle 1-loop Feynman diagram in 4d). 

\item The $(\Z_{8,\tA})$-(gravity)$^2$ anomaly captured by a perturbative anomaly (i.e., a triangle 1-loop Feynman diagram in 4d). 
The gravity part is due to the diffeomorphism of the background geometry.

\item The $(\Z_{8,\tA})$-(SU(2)$_R$)$^2$ anomaly captured by a perturbative anomaly (i.e., a triangle 1-loop Feynman diagram in 4d). 

\end{enumerate}

Ref.~\cite{2018arXiv180609592C} explains the two interesting mixed 't Hooft higher anomalies involving 1-form symmetry,
the Type I \eq{eq:type1} and Type II \eq{eq:type2} anomalies earlier.

\begin{enumerate}
  \setcounter{enumi}{4}
\item  Type I higher anomaly:
 mixing  
between
 the 1-form electric center symmetry ($\Z_{2,[1]}^e$)
and
 the 0-form discrete axial symmetry ($\Z_{2 N_c N_f}$ = $\Z_{8}$). 
 We can write \eq{eq:type1} as
\bea
&&\e^{\ii \frac{k\pi}{2} \int A \cup \mathcal{P}_2(B_2)} \nn \\
&&=\e^{\ii \frac{k \pi}{2} \int A \cup (B_2\cup B_2+B_2\cup_1\delta B_2)} \nn\\
&&=\e^{\ii \frac{k\pi}{2} \int A \cup (B_2\cup B_2+B_2 (2 \Sq^1  B_2))},
\eea
see \cite{W2} for introducing the cup products, higher cup products and the Steenrod square Sq.

\item  Type II higher anomaly:
mixing between
 the 1-form center symmetry (denoted as $\Z_{2,[1]}$)
and the background gravity (or the curved spacetime geometry) in \eq{eq:type2}.

\end{enumerate}

The UV theory as an adjoint QCD$_4$ has all of the above 't Hooft anomalies, captured also by a particular 5d cobordism invariant, in \eq{eq:Spin-cob} and \eq{eq:Pin-cob}. 

Following our Introduction,
in the next \Sec{sec:higher-sym-ext}, we formulate the higher {symmetry-extension} generalizing \cite{Wang2017loc170506728},
and successfully construct a 4d symmetric anomalous TQFT for Type II anomaly \eq{eq:type2}.
But we will soon show an obstruction to construct symmetric TQFT for the Type I anomaly \eq{eq:type1}.

\section{Higher Symmetry-Extension}
\label{sec:higher-sym-ext}

\subsection{Summary of Ordinary Symmetry-Extension 
}

\label{sec:WWW}

Ref.~\cite{Wang2017loc170506728} sets up the symmetry-extension problem as follows. 
Consider the $d$d SPTs protected by an internal symmetry group $G$, 
whose boundary theory has $(d-1)$d 't Hooft anomaly in $G$. 
There are three different ways to phrase the question asked by \cite{Wang2017loc170506728}, but their underlying meanings are the same:
%


\begin{enumerate}[label=\textcolor{blue}{Q\arabic*.}, ref={Q\arabic*},leftmargin=*]
\item \label{q1}
{\bf Condensed matter statement}: Can we find a total group $H$ such that $G$ is its quotient group,
and such that the $G$-SPTs becomes a trivial gapped vacua in $H$? More precisely, 
there is a local unitary transformation preserving the symmetry $H$ (but breaking the symmetry $G$),
such that when the $G$-SPTs is viewed as an $H$-SPTs, it can be deformed to 
a trivial gapped insulator in $H$ via a local unitary transformation, without breaking $H$ 
and without any phase transition.\footnote{This procedure has been demonstrated explicitly in a many body quantum system recently in Ref.~\cite{Prakash2018ugo1804.11236},  
which constructs an explicit path in the enlarged $H$-symmetric quantum Hilbert space.} 
\item \label{q2} 
{\bf QFT or high energy particle physics statement}:
Given a $(d-1)$d 't Hooft anomaly in $G$, can we find an enlarged group $H$,
with a total group $H$ having $G$ as its quotient group,
such that the 't Hooft anomaly in $G$ becomes \emph{anomaly-free} in $H$? (i.e., the $G$-anomaly  becomes trivial in $H$.)

\item \label{q3} 
{\bf Mathematical and algebraic topology statement}:
Given a $d$d topological term of a group $G$, here the topological term can be:
\begin{itemize}
\item the $d$d cocycle for a $d$-th cohomology group $\H^d(\B G,\U(1))$  in a group cohomology theory.
\item the $d$d co/bordism invariant for a $d$-th cobordism group $\Omega^d_{-}(\B G,\U(1))$ or bordism group $\Omega_d^{-}(\B G)$ or  bordism group, in a cobordism theory;\footnote{Here the $-$ can be chosen as co/bordism  with different structures such as special/orthogonal $\SO$/$\tO$, spin/pin, or $\Spin$/$\Pin^{\pm}$ structures.
\label{ft:-}}
\end{itemize}
can we find an extended group $H$ with $G$ its quotient group, via a short exact sequence 
\bea
1 \to K \to H\to G\to 1,
\eea 
such that the topological term of a group $G$ can be pulled back to a trivial topological term of a group $H$?
\end{enumerate}

Suppose the above answer is positive, 
and suppose that $G$, $H$ and $K$ are finite groups, then Ref.~\cite{Wang2017loc170506728} shows, 
valid for both the lattice Hamiltonian and the path integral construction, 
that the $G$-SPTs in $d$d can allow:\\ 
$\bullet$ $H$-symmetry extended gapped boundary in any spacetime dimension $d \geq 2$, \\
$\bullet$  $G$-symmetry preserving and topological $K$-gauge theory gapped boundary:
Topological emergent $K$-gauge theory with preserving global symmetry $G$ on a bulk $d \geq 3$.\\ 

Ref.~\cite{Wang2017loc170506728} addresses the above questions  \ref{q1}, \ref{q2} and  \ref{q3},
 by proving that at least for a finite group $G$ (with $G$ a unitary symmetry group or anti-unitary symmetry group involving time-reversal symmetry),
 by the following positive answers, with the \emph{always-existences} on the validity of the symmetric gapped boundary construction:

\begin{enumerate}[label=\textcolor{blue}{A\arabic*.}, ref={A\arabic*},leftmargin=*]
\item \label{a1}
{For any bosonic 
SPT state with a finite onsite symmetry group $G$, including both unitary and anti-unitary symmetry, there \emph{always exists} an 
$H$-symmetry-extended (or $G$-symmetry-preserving) gapped boundary 
via a nontrivial group extension by a finite $K$, given the bulk spacetime dimension $d \geq 2$}.

\item \label{a2}
{For any $G$-anomaly in $(d-1)$d given by a cocycle $\nu_d^G \in \H^d(\B G,\U(1))$ of group cohomology of a finite group $G$, 
there \emph{always exists} a pull back to a finite group $H$ via 
a certain
group extension $1 \rightarrow K \rightarrow  H  \overset{r}{\rightarrow} G \rightarrow 1$, extended by a finite $K$, 
such that $G$-anomaly becomes $H$-anomaly free, given the dimension $d \geq 2$}

\item \label{a3}
{For any $G$-cocycle $\nu_d^G \in \H^d(\B G,\U(1))$ of a finite group $G$, 
there \emph{always exists} a
pull back to a finite group $H$ via 
a certain short exact sequence of a 
group extension $1 \rightarrow K \rightarrow  H  \overset{r}{\rightarrow} G \rightarrow 1$ by a finite $K$, 
such that 
$${r}^* \nu_d^{G}= \nu^H_d =\delta  \mu^H_{d-1} \in \H^d(H,\U(1)).$$ 
Here $r$ is the pullback operation, and $\delta$ is the coboundary operation.
Namely,
a $G$-cocycle becomes a $H$-coboundary, which splits to a one-lower dimensional $H$-cochains $\mu^H_{d-1}$, 
given the dimension $d \geq 2$.}

\end{enumerate}

The proof of \cite{Wang2017loc170506728} has also been verified later by \cite{Tachikawa2017gyf}.
The related constructions similar to \cite{Wang2017loc170506728} are explored also in specific cases or from different perspectives
in \cite{Kapustin1404.3230, Cheng:2017kew1606.08482}.

\subsection{Higher Symmetry Generalization}

Now we generalizes the approach in \cite{Wang2017loc170506728}.
The short exact sequence of a 
group extension $1 \rightarrow K \rightarrow  H  \overset{r}{\rightarrow} G \rightarrow 1$ extended by a finite $K$
given in \cite{Wang2017loc170506728}
also implies an induced fiber sequence from the fibration 
\bea \B K\to \B H \to \B G,
\eea
where all $G,K$ and $H$ are finite groups of 0-form symmetry such that the $G$-SPTs protected by a finite group $G$ becomes trivial $H$-SPTs by pulling pack $G$ to $H$,
under the above criteria 
 \ref{a1}, \ref{a2} and  \ref{a3}.

{
We consider the higher symmetry-extension problem. 
A simpler example is 
$$
{\B K_{[0]} \times \B^2 K_{[1]}}\to \B\mathbb{H}\to\B\mathbb{G},
$$
where $K_{[0]}$ is an extension from a normal 0-form symmetry $K_{[0]}$,
while $K_{[1]}$ is an extension from a less familiar and more exotic 1-form symmetry $K_{[1]}$.
However, our goal is more ambitious to check a more general fibration
\bea \label{eq:2-group-fiber}
{\B K_{[0]} \ltimes \B^2 K_{[1]}}\to \B\mathbb{H}\to\B\mathbb{G}
\eea 
where $\mathbb{G}$ and $\mathbb{H}$ are 2-groups,
$K_{[0]}$ and $K_{[1]}$ are finite abelian groups of 0-form symmetry and 1-form symmetry respectively such that the 
higher-$\mathbb{G}$-SPTs protected by a 2-group $\mathbb{G}$ becomes the trivial 
higher-$\mathbb{H}$-SPTs by pulling back 
$\mathbb{G}$ to $\mathbb{H}$.\footnote{For the related physics topics on higher group symmetries and higher SPTs, the readers can find from the recent developments
\cite{Cordova:2018cvg2group, Benini:2018reh, Delcamp2018wlb1802.10104, Zhu2018kzd1808.09394, Wen2018zux1812.02517, Delcamp2019fdp1901.02249} and References therein.}
{
Here $
{\B K_{[0]} \ltimes \B^2 K_{[1]}}$ is the total space $\B\mathbb{K}$ of the fibration
\bea
\xymatrix{\B^2K_{[1]}\ar[r]&\B\mathbb{K}\ar[d]\\
&\B K_{[0]}.}
\eea
}
}

Similar to questions in \ref{q1}, \ref{q2} and  \ref{q3}  of \Sec{sec:WWW}, we ask a set of generalized questions:
\begin{enumerate}[label=\textcolor{blue}{Q\arabic*.}, ref={Q\arabic*}, leftmargin=*]
  \setcounter{enumi}{3}
\item \label{q4}
{\bf Condensed matter statement}: Can we find a total 2-group $\mathbb H$ as a total space such that $\B\mathbb G$ is $\B\mathbb H$'s orbit (or base space),
and such that the $\mathbb G$-SPTs becomes a trivial gapped vacua in $\mathbb H$? More precisely, 
there is a local unitary transformation preserving the symmetry $\mathbb H$ (but breaking the symmetry $\mathbb G$),
such that when the $\mathbb G$-SPTs is viewed as an $\mathbb H$-SPTs, it can be deformed to 
a trivial gapped insulator in $\mathbb H$ via a \emph{local} unitary transformation (note that the locality also need to be generalized to higher dimensional extended object such as a line instead of just a point, due to the 2-group structure), 
without breaking $\mathbb H$ 
and without any phase transition in the enlarged $\mathbb H$-symmetric quantum Hilbert space. 
\item \label{q5} 
{\bf QFT or high energy particle physics statement}:
Given a $(d-1)$d 't Hooft anomaly in a higher group $\mathbb G$, can we find an enlarged group $\mathbb H$,
with a total group $\mathbb H$ obeying \eq{eq:2-group-fiber},
such that the 't Hooft anomaly in $\mathbb G$ becomes \emph{anomaly-free} in $\mathbb H$? (i.e., the $\mathbb G$-anomaly becomes trivial in $\mathbb H$.)

\item \label{q6} 
{\bf Mathematical and algebraic topology statement}:
Given a $d$d topological term of a higher group $\mathbb G$, here the topological term can be:
\begin{itemize}
\item the $d$d cocycle for a $d$-th cohomology group $\H^d(\B \mathbb G,\U(1))$  in a higher group cohomology theory.
\item the $d$d co/bordism invariant for a $d$-th cobordism group $\Omega^d_{-}(\B \mathbb G,\U(1))$ or bordism group $\Omega_d^{-}(\B \mathbb G)$ or  bordism group, in a cobordism theory;\footnote{Here the ``$-$'' follows the earlier footnote \ref{ft:-}.}
\end{itemize}
can we find an extended group $\mathbb H$ obeying \eq{eq:2-group-fiber}
such that the topological term of a group $\mathbb G$ can be pulled back to a trivial topological term of a group $\mathbb H$?
\end{enumerate}

In the next two subsections, we implement the strategy \eq{eq:2-group-fiber} by asking the 
questions in \ref{q4}, \ref{q5} and  \ref{q6},
for the two examples:
{Type I anomaly/topo.~invariant} in \eq{eq:type1},
and
{Type II anomaly/topo.~invariant} in \eq{eq:type2}.

We relegate more formal and mathematical details of the calculation of the above two subsections into Appendices  \ref{sec:cobordism-TQFT}, \ref{sec:proof}, and \ref{sec:P(B)}.

\subsection{Saturate {Type II anomaly}: Symmetric TQFTs}

We first try to do higher symmetry extension to trivialize 4d Type II higher anomaly (given by a 5d topological invariant) \eq {eq:type2}
$$
\e^{\ii {\pi}  \int w_2(TM) \Sq^1 B_2}=\e^{\ii {\pi}  \int w_3(TM) B_2}.
$$ 
We have found that \eq {eq:type2} is a topological invariant in $d=5$, for:\\
$\bullet$ $\H^d(\B^2 \Z_2,\U(1))$ group cohomology of a higher classifying space finite group, as well as\\
$\bullet$  $\Omega_d^{\SO}(\B^2 \Z_2)$ cobordism group of a higher classifying space finite group.
Below we can either use the group cohomology or the cobordism group viewpoint to understand the
trivialization of 4d Type II higher anomaly.

\begin{enumerate}[leftmargin=*] 

\item
The first way to trivialize this 4d Type II higher anomaly is extending the spacetime symmetry from special orthogonal group
$\SO(d)= \Spin(d)/\Z_2^F$ to $\Spin(d)$:
\bea \label{Type-II-extend-1}
\B \Z_2 \to \B \Spin(d) \times \B^2 \Z_2 \to \B \SO(d) \times \B^2 \Z_2.
\eea
This extension works since $w_2(TM) =0$ vanishes on Spin manifold.
Thus,  \eq {eq:type2} is trivialized once we pull back  \eq {eq:type2} into
$\B \Spin(d) \times \B^2 \Z_2$. 
According to the interpretation in
\Sec{sec:WWW} and Ref.~\cite{Wang2017loc170506728}, 
the fibration $\B \Z_2$ contains an emergent 0-form global symmetry
which is anomaly-free and can be dynamically gauged.
Indeed, the natural way to interpret the \eq{Type-II-extend-1} 
as the generalized construction of \cite{Wang2017loc170506728}
is that
there is an emergent 1-form $\Z_2$ gauge theory 
(dynamically gauged from emergent 0-form global symmetry $\B \Z_2$),
such that the $\Z_2$ gauge theory has additional emergent \emph{fermionic} particle excitations
due to the emergent spin structure (the Spin$(d)$ in the total space in \eq{Type-II-extend-1}).
In terms of the full 4d symmetric TQFT saturating the higher 't Hooft anomaly (coupling to the 5d higher SPTs), we can write
the involved QFT sectors into a partition function, which looks like the following \emph{locally}:
\begin{widetext}
\bea \label{eq:TQFT-1}
\underbrace{\e^{\ii {\pi}  \int_{M^5} w_2(TM) \Sq^1 B_2} }_{\text{5d higher SPTs (4d higher anomaly)}}
\cdot 
\underbrace{\sum_{{a \in C^1( (\partial M)^4, \Z_2),}\atop{b \in C^2( (\partial M)^4, \Z_2)}} 
\exp( \ii  2 \pi  \int_{(\partial M)^4} \frac{1}{2} (b \delta a 
 ) +\dots )}_{\text{locally a 4d $\Z_2$-TQFT with emergent fermions and spin-structure}}.
\eea
Here $a$ is the $\Z_2$-valued 1-form gauge field (the standard notation as the 1-cochain in $C^1$),
 $b$ is the $\Z_2$-valued 2-form gauge field (the standard notation as the 2-cochain in $C^2$),
the $\delta$ is the coboundary operator here $\delta = 2 \Sq^1$, and we use the cup product $\cup$.
See also our previous explanations around \eq{eq:type2} for notations.
The $\dots$ are additional coupling terms between dynamical gauge fields and background fields.
The $\dots$ also include additional sectors from the UV adjoint QCD$_4$ from \eq{eq:ZUV}, in order to saturate the other anomalies.
Note that the similar emergent dynamical spin structure with $\Z_2$ gauge field has been studied in \Ref{Wang:2018qoyWWW}.
The important thing is that the 1-form gauge field $a$ can be regarded as the difference between two spin-structures,
while the gauge field $a$ becomes dynamical.

Moreover, we can write the extension of \eq{Type-II-extend-1} in terms of the full symmetry \eq{eq:Spin-sym}:
\bea  \label{Type-II-extend-all}
\B \Z_2 \to \B ({\Spin \times \big(\frac{\SU(2) \times \Z_{8,\tA}}{\Z_2^F} \big))  \times  \B^2 \Z_{2,[1]}^e} \to 
\B({\Spin \times_{{\Z_2^F}} \big(\frac{\SU(2) \times \Z_{8,\tA}}{\Z_2^F} \big)) \times \B^2  \Z_{2,[1]}^e},
\eea
while the physical interpretation remains the same as  \eq{Type-II-extend-1} and \eq{eq:TQFT-1}.
\end{widetext}

\item
The second way to trivialize this 4d Type II higher anomaly is extending the 1-form symmetry:
\bea \label{Type-II-extend-2}
\B^2 \Z_2 \to \B \SO(d) \times \B^2 \Z_4 \to \B \SO(d) \times \B^2 \Z_2.
\eea
This way works since $B$ is pulled back to $\tilde B\in\H^2(\B^2\Z_4,\Z_2)$, and $\Sq^1\tilde B=0$ (see Appendix \ref{sec:third}).

According to the interpretation in
\Sec{sec:WWW} and Ref.~\cite{Wang2017loc170506728}, 
the fibration $\B^2 \Z_2$ is associated to an emergent 1-form global symmetry $\Z_{2,[1]}$
which is anomaly-free and can be dynamically gauged.
Indeed, the natural way to interpret the \eq{Type-II-extend-2} 
as the generalized construction of \cite{Wang2017loc170506728}
is that
there is an emergent 2-form $\Z_2$ gauge theory 
(dynamically gauged from emergent 1-form global symmetry $\B \Z_2$) with a 2-form gauge field $b'$.
The original 1-form $\Z_{2,[1]}^e$-symmetry acts projectively on the emergent 2-form $\Z_2$ gauge theory,
but the extended 1-form $\Z_{4,[1]}^e$-symmetry acts on it faithfully.  
\begin{widetext}
{
We can write
the involved QFT sectors into a following partition function, which looks like the following \emph{locally}:
\bea \label{eq:TQFT-2}
\underbrace{\e^{\ii {\pi}  \int_{M^5} w_2(TM) \Sq^1 B_2} }_{\text{5d higher SPTs (4d higher anomaly)}}
\cdot 
\underbrace{\sum_{{a' \in C^1( (\partial M)^4, \Z_2),}\atop{b' \in C^2( (\partial M)^4, \Z_2)}} 
\exp( \ii  2 \pi  \int_{(\partial M)^4} \frac{1}{2} (a' \delta b' 
) +\dots )}_{\text{locally a 4d $\Z_2$-TQFT, on which the 1-form $\Z_{2,[1]}^e$-symmetry acts projectively}}.
\eea
}
Here $b'$ is the $\Z_2$-valued 2-form gauge field (the standard notation as the 2-cochain in $C^2$),
$a'$ is the $\Z_2$-valued 1-form gauge field (the standard notation as the 1-cochain in $C^1$),
while other notations are explained around \eq{eq:type2} and \eq{eq:TQFT-2}.
The $\dots$ are additional coupling terms between dynamical gauge fields and background fields.
The $\dots$ also include additional sectors from the UV adjoint QCD$_4$ from \eq{eq:ZUV}, in order to saturate the other anomalies.
We can also write the extension of \eq{Type-II-extend-2} in terms of the full symmetry \eq{eq:Spin-sym}:
\bea 
\B^2 \Z_{2,[1]} \to \B({\Spin \times_{{\Z_2^F}} \big(\frac{\SU(2) \times \Z_{8,\tA}}{\Z_2^F} \big))  \times  \B^2 \Z_{4,[1]}^e} \to 
\B({\Spin \times_{{\Z_2^F}} \big(\frac{\SU(2) \times \Z_{8,\tA}}{\Z_2^F} \big)) \times \B^2  \Z_{2,[1]}^e},
\eea
while the physical interpretation remains the same as  \eq{Type-II-extend-2}.
\end{widetext}

\end{enumerate}

\subsection{
Saturate {Type I anomaly}: Obstruction}

We now try to do higher symmetry extension to trivialize 4d Type I higher anomaly (given by a 5d topological invariant of higher SPTs) \eq {eq:type1}
$$
\e^{\ii \frac{k\pi}{2} \int A \cup \mathcal{P}_2(B_2)}
=\e^{\ii \frac{k \pi}{2} \int A \cup (B_2\cup B_2+B_2\cup_1\delta B_2)}
$$ 
Below we show that
\begin{enumerate}[leftmargin=*] 
\item
When $k=2 \in \Z_4$, the Type I anomaly \eq {eq:type1} can be trivialized,
thanks to the fact that we can rewrite \eq {eq:type1} as
\bea
&&\e^{\ii \pi \int A \cup \mathcal{P}_2(B_2)}\nn\\
&=&\e^{\ii  \pi \int A \cup (B_2\cup B_2+B_2\cup_1\delta B_2)}\nn\\
&=&\e^{\ii  \pi \int A \cup (B_2\cup B_2+2B_2\cup_1\Sq^1 B_2)}\nn\\
&=&\e^{\ii  \pi \int A \cup (B_2\cup B_2)}\nn\\
&=&\e^{\ii  \pi \int \Sq^2(A \cup B_2)}\nn\\
&=&\e^{\ii  \pi \int (w_2(TM)+w_1(TM)^2)(A \cup B_2)},
\eea
where we have used the fact that $\Sq^1\tilde A=0$ where $\tilde A=A\mod2$, and the Wu formula.
See also useful information in \cite{W2}.

So when $k=2 \in \Z_4$, if we extend the global symmetry by
\bea \label{eq:k2TypeII}
&&\B\Z_2\to\B(\Spin\times(\SU(2)\times_{\Z_2}\Z_8))\times\B^2\Z_2 \to \nn\\
&&
\B(\Spin\times_{\Z_2}(\SU(2)\times_{\Z_2}\Z_8))\times\B^2\Z_2,
\eea
then the Type I anomaly \eq {eq:type1} vanishes.
This extension works since $w_1(TM) =w_2(TM) =0$ vanishes on Spin manifolds.
Thus,  \eq {eq:type1} is trivialized once we pull back  \eq {eq:type1} into
$\B(\Spin\times(\SU(2)\times_{\Z_2}\Z_8))\times\B^2\Z_2 $. 

\item
When $k=1,3$ $\in \Z_4$, or $k$ odd, the Type I anomaly \eq {eq:type1} cannot be trivialized by extensions.

We have tried three approaches, which we relegate the details in Appendix \ref{subsec:pullback} while we summarize the physics story and implication here.

\begin{itemize}[leftmargin=*] 
\item
The first approach (Appendix \ref{subsec:1})
 is a breaking case since we set $B$ to be zero.
Physically this means that in order to saturate the 't Hooft anomaly, 
we can break 1-form $\Z_2$-symmetry to nothing.
In comparison, this 1-form $\Z_2$-symmetry breaking 
is a different scenario from \cite{2018arXiv180609592C,Bi:2018xvr}.
 
 \item
In the second approach (details and notations explained in Appendix \ref{subsec:2}), we define $\mathbb{G}$ to be a group which sits in a homotopy pullback square
\bea
\xymatrix{
\B \mathbb{G}\ar[d]\ar[rr]&&\B ^2\Z_2\ar[d]^{x_2}\\
\B (\Spin\times\SU(2)\times\Z_8)\ar[r]^-{j_3}&\B \Z_8\ar[r]^{\tilde{b}}&\B ^2\Z_2.}
\eea
Hence we have a fiber sequence
\be
\B\Z_2\to \B\mathbb{G}\to \B(\Spin\times\SU(2)\times\Z_8)\times \B^2\Z_2\to \B^2\Z_2.
\ee

In this case, $B_2=B$ is identified with $\beta_{(2,8)}A$ where $A\in\H^1(\B\Z_8,\Z_8)$, and $\Sq^1B=0$, but $\tilde A\cup\frac{B^2}{2}$ is still not trivialized.
This case is also a breaking case, 
{since $B$ is locked with $A$.
In physics, the locking between two probed background fields means that the global symmetry between two sectors are locked together, thus which results in 
global symmetry breaking.}

Physically this means that in order to saturate the 't Hooft anomaly, 
we still need to break symmetry in some way.

\item
In the third approach, we extend both the 0-form symmetry and the 1-form symmetry:
\bea
&&\B\Z_2 \times \B^2\Z_2\to \B(\Spin\times(\SU(2)\times_{\Z_2} \Z_8)) \times  \B^2 \Z_4\to \nn\\
&&\B(\frac{\Spin\times(\SU(2)\times_{\Z_2} \Z_8)}{{\Z_2^F}})\times \B^2\Z_2.
\eea
But in this case, $\tilde A\cup\frac{B^2}{2}$ is still not, and cannot be, trivialized.
\end{itemize}
\end{enumerate}

\begin{widetext}

In summary, 
we finally conclude that when $k$ is odd, $k=1,3 \in \Z_4$, the Type I anomaly \eq {eq:type1} cannot be trivialized by extensions and give a proof in Appendix \ref{sec:proof}.
In comparison, Ref.~\cite{Bi:2018xvr} proposes a full symmetry-preserving TQFT  
different all of our scenarios above, which contradicts to our proof in Appendix \ref{sec:proof}.

\subsection{
Saturate both {Type I (for even class in $\Z_4$) and II anomalies}}

When $k=2$, such that the Type I anomaly survives as only a $\Z_2$ subclass (even $k$) in the original $k \in \Z_4$ class (of $k A \cup \mathcal{P}_2(B_2)$),
however, we can actually trivialize the $\Z_2$ subclass of Type I anomaly and the full Type II anomaly together via the fibration: 
\bea   \label{eq:k2TypeI-II-1}
\B \Z_2 \to \B ({\Spin \times \big(\frac{\SU(2) \times \Z_{8,\tA}}{\Z_2^F} \big))  \times  \B^2 \Z_{2,[1]}^e} \to 
\B({\Spin \times_{{\Z_2^F}} \big(\frac{\SU(2) \times \Z_{8,\tA}}{\Z_2^F} \big)) \times \B^2  \Z_{2,[1]}^e}.
\eea
The above is achieved by combining both \eq{Type-II-extend-1} and \eq{eq:k2TypeII} into \eq{Type-II-extend-all}.
Since we only care $k=2$, this also means that the $\Z_{8,\tA}$-symmetry only needs to be survived as a $\Z_{4,\tA}$-symmetry.
Physically this means that $\Z_{8,\tA}$-symmetry can be spontaneously broken down to a $\Z_{4,\tA}$-symmetry.
Thus the \eq{eq:k2TypeI-II-1} really implies a fibration of a smaller symmetry (e.g. a smaller classifying space) as:
\bea   \label{eq:k2TypeI-II-2}
\B \Z_2 \to \B ({\Spin \times \big(\frac{\SU(2) \times \Z_{4,\tA}}{\Z_2^F} \big))  \times  \B^2 \Z_{2,[1]}^e} \to 
\B({\Spin \times_{{\Z_2^F}} \big(\frac{\SU(2) \times \Z_{4,\tA}}{\Z_2^F} \big)) \times \B^2  \Z_{2,[1]}^e}.
\eea
For such a 4d TQFT preserving a $(\frac{\SU(2) \times \Z_{4,\tA}}{\Z_2^F})$-chiral symmetry and 1-form $\Z_{2,[1]}^e$-symmetry (from UV adjoint QCD$_4$),
saturating the higher 't Hooft anomaly (coupling to the 5d higher SPTs), we can write
the involved QFT sectors into a partition function, which looks like the following \emph{locally}:
\bea \label{eq:TQFT-all}
\underbrace{\e^{\ii \pi \int A \cup (B_2\cup B_2)} \cdot \e^{\ii {\pi}  \int_{M^5} w_2(TM) \Sq^1 B_2} }_{\text{5d higher SPTs (4d higher anomaly)}}
\cdot 
\underbrace{\sum_{{a \in C^1( (\partial M)^4, \Z_2),}\atop{b \in C^2( (\partial M)^4, \Z_2)}} 
\exp( \ii  2 \pi  \int_{(\partial M)^4} \frac{1}{2} (b \delta a 
 ) +\dots )}_{\text{locally a 4d $\Z_2$-TQFT with emergent fermions and spin-structure}},
\eea
again the 1-form gauge field $a$ can be regarded as the difference between two spin-structures;
the 1-form emergent dynamical $\Z_2$ gauge field $a$ is associated to 
a dynamical spin structure (similar to a situation in \Ref{Wang:2018qoyWWW}).

We note that the $\dots$ terms can involve additional 't Hooft anomaly cancellation for the UV's adjoint QCD$_4$, such as the gapless sector proposed in
\cite{Anber2018tcj1805.12290,2018arXiv180609592C,Bi:2018xvr, SWW}. 
Besides, the $\dots$ terms also involve the coupling terms 
between dynamical gauge fields and background fields, so that the full partition function can be made gauge-invariant.
Although \eq{eq:k2TypeI-II-2} already suggests a formation definition of TQFTs 
(based on the extension construction of bulk-boundary coupled TQFTs, see \cite{Wang2017loc170506728} and related constructions 
in \cite{2018PTEP1801.05416, Guo2018vij1812.11959}),
it may be worthwhile to formulate a cochain or continuum TQFT description following  \cite{2018PTEP1801.05416, Guo2018vij1812.11959} 
--- which we leave them for future work.
It may also be worthwhile to give a continuum 4d TQFT formulation for the higher-form gauge theory analogous to Dijkgraaf-Witten \cite{Dijkgraaf:1989pz} like gauge theory,
similar to the continuum TQFT formulation given in \cite{Putrov2016qdo1612.09298, Delcamp2018wlb1802.10104}.
\end{widetext}

\subsection{Other Examples}

In our companion work \cite{Wan2018zql1812.11968}, 
we consider similar trivialization problem for 5d topological invariants of 4d Yang-Mills SU(N) gauge theory (in particular at $\theta=\pi$) anomaly. 
The successful trivialization by the pulling back via a finite group extension suggests 
that the low energy fate of these 4d Yang-Mills SU(N) gauge theories can  also be
a fully symmetry-preserving TQFT with the full 't Hooft anomaly matched.

The fully symmetry-preserving TQFT means that there is no 0-form global symmetry (e.g. time-reversal symmetry) breaking, nor the 1-form center symmetry breaking
(thus it is in a confined phase in the usual definition, where the charged Wilson loop is
tension-ful and has an area law); but there is an additional emergent discrete gauge theory as a TQFT associated to the finite group extension. 

In recent work, \Ref{Wan2019oyrWWZ1904.00994}
studies the related  fully symmetry-preserving TQFTs of various 
4d Yang-Mills SU(2) gauge theories where its SU(2)-fundamental Wilson lines can be Kramers singlet/doublet and bosonic/fermionic.
\Ref{Wan2019oyrWWZ1904.00994} uses the higher-symmetry extension method to suggest where the 4d UV Yang-Mills SU(2) gauge theories can flow to
candidate fully symmetry-preserving TQFTs
as new reasonable IR fates.

{We find that many other examples of 5d topological invariants of 4d Yang-Mills anomaly can be trivialized by extending the 0-form symmetry and 1-form symmetry.}
Hence, the higher symmetry-extension generalization \Ref{Wang2017loc170506728}
is powerful enough to trivialize a lot of other higher bosonic types of anomalies thus to construct exotic fully-symmetric anomalous TQFTs, 
although 
it gives an obstruction to saturate the Type I anomaly at an odd $k$ while preserving the full symmetry.


\section{Conclusion}
\label{sec:conclude}

We conclude by summarizing the implications of the higher-symmetry extension construction of TQFTs
on the low energy dynamics of QCD$_4$. Then we comment about the constraints on the 
deconfined quantum critical phenomena, or so called the deconfined quantum critical point (dQCP) \cite{dQCP2004}, in 3+1 spacetime dimensions \cite{Bi:2018xvr}.

\subsection{The Fate of the Dynamics of QCD$_4$}

\subsubsection{Possible Fates of the Dynamics of Fundamental QCD$_4$ with $N_f$ Dirac fermions}

\label{sec:fate-fund}

First, we recall the possible fates of the dynamics of QCD$_4$ with $N_f$ Dirac fermions in fundamental representations of SU($N_c$).
The conventional wisdom teaches us that the phase structure of dynamics of QCD$_4$ via tuning $N_f$ (with a fix $N_c$), shown in \Fig{Fig-1:FundQCD}, is that:\\

\noindent
$\bullet$ At lower $N_f$, there should be a confinement (IR confinement) and chiral symmetry breaking (IR ChSB).\\
$\bullet$ At larger $N_f$, 
there is a range of $N_f$, such that at IR, the QFT flows to 
an interacting conformal field theory (``CFT''),
this is known as the range of  \emph{conformal window} phenomena studied by Bank-Zaks \cite{Banks1981nnZaks} and others.\\
$\bullet$ Let $N_f^{\text{asym.free}}=\frac{11}{2}N_c$,
when $N_f < N_f^{\text{asym.free}}$, the UV theory is weak coupling known as the asymptotic freedom (or UV free) \cite{Gross1973idWilczekPRL, Politzer1973fxPRL}.
When $N_f > N_f^{\text{asym.free}}$,  the UV theory becomes strongly coupled while the  coupling $g$ flows weak at IR, at least perturbatively.


\subsubsection{Possible Fates of the Dynamics of Adjoint QCD$_4$ with $N_f$ Weyl fermions}
\label{sec:fate-adjoint}

Now we organize the possible fates of the dynamics of QCD$_4$ with $N_f$ Weyl fermions in adjoint representations of SU($N_c$).
The possible phase structure of dynamics of QCD$_4$ via tuning $N_f$ (with a fix $N_c$) is shown in \Fig{Fig-2:AdjointQCD}. 
We remark that the candidate adjoint phases are summarized very elegantly in \cite{2018arXiv180609592C}, we recap into a concise \Fig{Fig-2:AdjointQCD},
while also list down the related 
Scenario \ref{s1},  \ref{s2},   \ref{s3}, and  \ref{s4}, from Ref.~\cite{2018arXiv180609592C, Bi:2018xvr},
and from the list summarized in \Sec{sec:fate-adjoint}.

\begin{widetext}

\onecolumngrid
\begin{figure}[!h]
\includegraphics{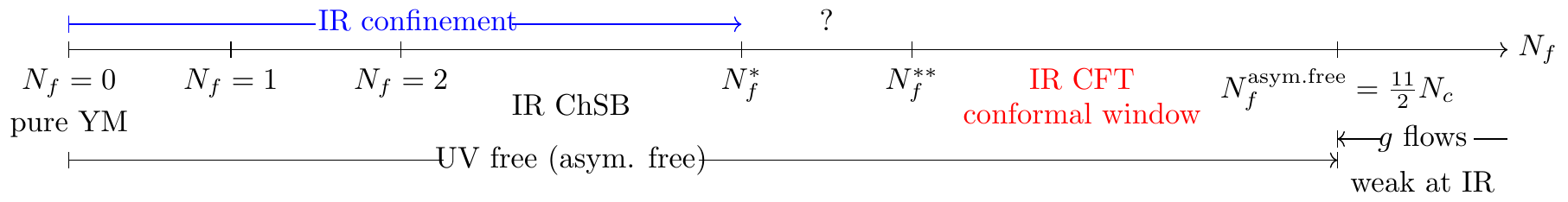}
\caption{Candidate phases of fundamental QCD$_4$ and their possible dynamical fates.
``ChSB'' means the ``chiral symmetry breaking phase.'' ``Pure YM'' means the pure Yang-Mills gauge theory with a SU($N_c$) gauge group.
``CFT'' means  conformal field theory.
``UV free'' or ``asym. free'' means the asymptotic free.
The question mark ``$?$'' means the detailed structure of the phase boundaries requires further studies. 
}
\label{Fig-1:FundQCD}
\end{figure}


\begin{figure}[!h]
\includegraphics{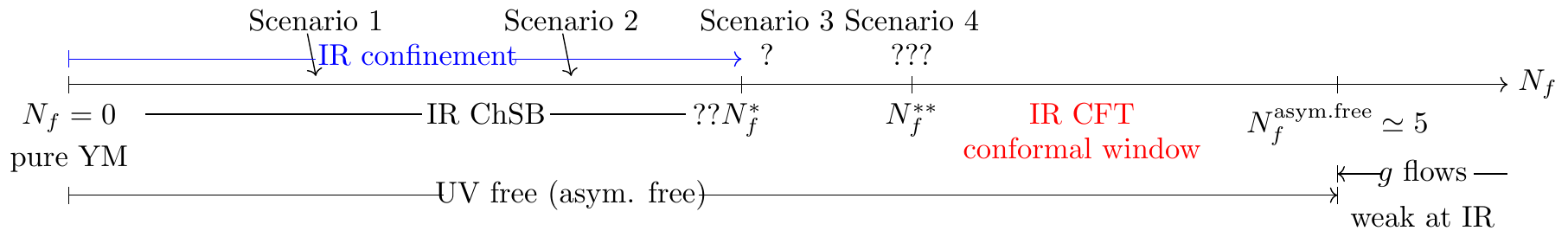}
\caption{
Candidate phases of adjoint QCD$_4$ with an SU(2) gauge group ($N_c=2$) and their possible dynamical fates.
``ChSB'' means the ``chiral symmetry breaking phase.''
The Scenario \ref{s1},  \ref{s2},   \ref{s3}, and  \ref{s4}
are from the list summarized in \Sec{sec:fate-adjoint}.  
The question marks ``$?$, $??$, and $???$'' means the detailed structure of the phase boundaries requires further studies. 
}
\label{Fig-2:AdjointQCD}
\end{figure}

\begin{figure}[!h]
\includegraphics{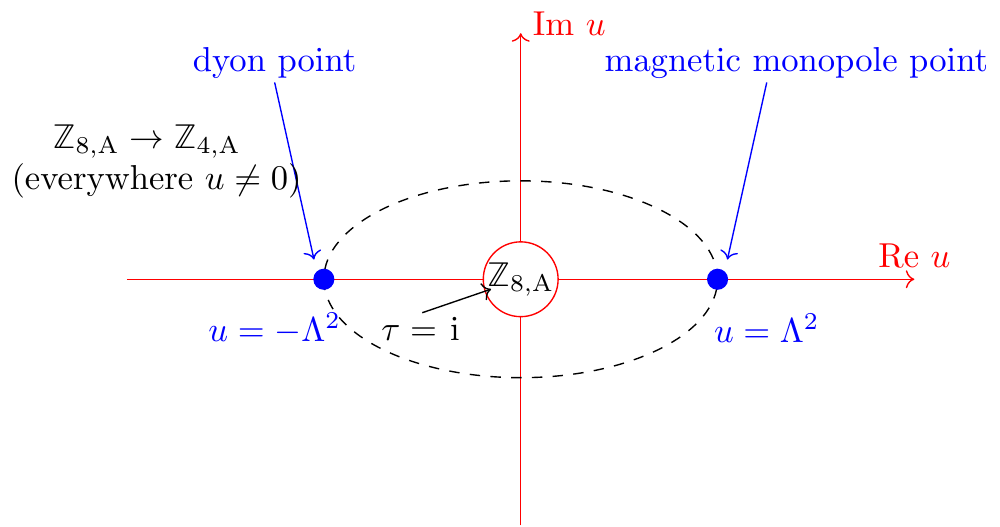}
\caption{
For the $\CN=2$ supersymmetric Yang-Mills theory (SYM) or Seiberg-Witten theory \cite{Seiberg1994rs9407087},
there is a moduli space of the supersymmetric vacua, labeled by the expectation value of 
the complex $u = \Re \; u + \ii \Im u \equiv \langle \Tr(\phi^2) \rangle \in \C $ for the 
$\CN=2$ chiral operator.
 The {$\tau \equiv \tau_{\U(1)} \equiv \frac{\theta}{2\pi}+ \frac{2\pi \ii}{e^2} =\ii$} is the special point at the self-dual value of the coupling {$\tau$}.
The coupling $\tau$ is for the Coulomb phase of U(1) gauge theory (the Coulomb branch for the moduli space of vacua).
 Even though the SYM is supersymmetric, \Ref{2018arXiv180609592C} enumerates the possible vacua by supersymmetry-breaking deformations.
 Let us relate the supersymmetric vacua to the adjoint QCD$_4$ vacua listed in the Scenarios \cite{2018arXiv180609592C}:  \\
 $\diamond$ The vacua of {magnetic monopole point} ({$u=\Lambda^2$}) and {dyon point} ({$u=-\Lambda^2$}) is related by the broken symmetry $\Z_{8, \tA}$ generator,
which gives rise a Scenario \ref{s1}.  \\
 $\diamond$ A generic vacua of $u \neq 0$, and $u \neq \pm \Lambda^2$ on the plane is related to a Scenario \ref{s2}.\\
 $\diamond$ The vacua of the $u=0$ with a self-dual coupling {$\tau_{}$} is related to a Scenario \ref{s4}.
}
\label{Fig-3:SW}
\end{figure}

\twocolumngrid

\end{widetext}


The conventional wisdom teaches us that the phase structure of dynamics of adjoint QCD$_4$ via tuning $N_f$ (with a fix $N_c$), shown in \Fig{Fig-2:AdjointQCD}, is that:\\

\noindent
\begin{itemize}[leftmargin=*] 
\item At $N_f=0$, it is a pure SU($N_c$) Yang-Mills gauge theory (say SU(2)), potentially with a $\theta$-term \eq{eq:theta}.
At $\theta=0$, the phase is a trivially gapped confined phase (IR confinement) with no SPT state.
However, at $\theta=\pi$, the phase has mixed higher anomalies  \cite{Gaiotto2017yupZoharTTT} and potentially newly found higher 't Hooft anomalies \cite{Wan2018zql1812.11968}. 
\item At $N_f=1$, it is a pure $\CN=1$ supersymmetric Yang-Mills gauge theory (SYM) \cite{Intriligator1995auSeiberg9509066}. 
Moreover,  there are $N_c$ supersymmetric breaking vacua due to gaugino condensation \cite{Witten1982dfNPBSUSY}, which breaks $\Z_{2 N_c}$ down to 
$\Z_{2}$ (simply $\Z_{2}^F$).
This $\CN=1$ SYM phase is also known to be confined through monopole condensation,
by embedding into a $\CN=2$ SYM theory with $N_c=2$ \cite{Seiberg1994rs9407087}.

\item At lower $N_f$, there should be a confinement (IR confinement) and chiral symmetry breaking (IR ChSB). 
\item At larger $N_f$, one expects again  
a range of $N_f$ with a range of  \emph{conformal window} phenomena of Bank-Zaks \cite{Banks1981nnZaks}.

\end{itemize}

To proceed further,
we recall that the UV internal global symmetry is
$\big(\frac{\SU(2) \times \Z_{8,\tA}}{\Z_2^F} \big) \times  \Z_{2,[1]}^e$.
Now we organize 
a list of {possible fates of the dynamics of adjoint QCD$_4$ with $N_f$ Weyl fermions}
proposed from \cite{2018arXiv180609592C, Bi:2018xvr}.
There are four {\bf scenarios}, summarized in Table \ref{table:adjoint} and below:

\begin{enumerate}[label=\textcolor{blue}{\arabic*.}, ref=\arabic*, leftmargin=*]

\item \label{s1}
The $N_c$ copies of (or more specifically here $N_c=2$) of 4d $\CP^1$ sigma model at 
low energy with spontaneous symmetry breaking Goldstone modes, proposed by \cite{2018arXiv180609592C}. 
Its global symmetry:
\bea
\tO(2) \times  \Z_{2,[1]}^e.
\eea
In summary, the scenario \ref{s1} has:
\bea
\text{``chiral symmetry breaking, and confinement.''}
\eea

To digest better about the target space of $\CP^1$ sigma model,
here we can consider the breaking of the 0-form symmetry group $G$ as the total space $E$ 
breaking to a smaller fiber $F$ (a subgroup or a normal subgroup, as the fiber or the stabilizer),
where the order parameter parametrizes the base manifold $B$ (the base space or the orbit). 
In short, we formally and mathematically write:
\bea
&&\begin{array}{ccl}
F&\stackrel{}{\hookrightarrow}& E\\
&&\downarrow{\scriptstyle }\\
&& B
\end{array},
\qquad
\begin{array}{ccl}
\text{stabilizer} &\stackrel{}{\hookrightarrow}& \text{total space}\\
&&\downarrow{\scriptstyle }\\
&& \text{orbit}
\end{array}.
\eea
Then we obtain a relation for the scenario \ref{s1}:
\bea
&&\begin{array}{ccl}
S^1=\U(1)_R&\stackrel{}{\hookrightarrow}& S^3=\SU(2)_R\\
&&\downarrow{\scriptstyle }\\
&& S^2=\CP^1
\end{array},
\eea
or more precisely a relation:
\bea
&&\begin{array}{ccl}
\tO(2)_R=\U(1) \rtimes \Z_2&\stackrel{}{\hookrightarrow}&\big(\frac{\SU(2)_R \times \Z_{8,\tA}}{\Z_2^F} \big)\\
&&\downarrow{\scriptstyle }\\
&& \mathbb{CP}^1  \rtimes \frac{\Z_{8,\tA}}{{\Z_2} \times\Z_2^F}.
\end{array}
\eea
The $ \mathbb{CP}^1  \rtimes \frac{\Z_{8,\tA}}{{\Z_2} \times\Z_2^F}$ has two copies of $\CP^1$ as the target space,
parametrizing the order parameter of the base manifold $B$ (the base space or the orbit).

\item \label{s2}
A free massless Dirac fermion (equivalently, two massless Weyl fermions, or two massless Majorana fermions)
and a $\Z_2$ discrete gauge theory as a 4d TQFT with a $\Z_4$ symmetry (spontaneously broken from the $\Z_8$ symmetry), proposed by \cite{2018arXiv180609592C}.
The IR symmetry is
\bea
\big(\frac{\SU(2) \times \Z_{4,\tA}}{\Z_2^F} \big) \times  \Z_{2,[1]}^e.
\eea
In summary, the scenario \ref{s2} has:
\be
\text{``chiral symmetry breaking $\Z_8 \to \Z_4$, and confinement.''}
\ee
However, as explained in \cite{2018arXiv180609592C},
there is an additional emergent 
new deconfined $\Z_2$-TQFT
with emergent new $\Z_{2,[1]}$ symmetries spontaneously broken.

\item \label{s3}
A free massless Dirac fermion (equivalently, two massless Weyl fermions, or two massless Majorana fermions)
and a 4d TQFT preserving the full $\Z_8$ symmetry, proposed by \cite{Bi:2018xvr}.
The two massless Weyl fermions actually have a U(2) continuous global symmetry.
The IR symmetry we focus is:
\bea
\big(\frac{\SU(2) \times \Z_{8,\tA}}{\Z_2^F} \big) \times  \Z_{2,[1]}^e.
\eea
In summary, the scenario \ref{s3} proposed that:
\be
\text{``chiral symmetry fully preserved, and confinement.''}
\ee

\item \label{s4}
A 4d U(1) gauge theory in Coulomb phase with a $\Z_{2 N_c N_f}$ = $\Z_{8}$ symmetry, proposed by \cite{2018arXiv180609592C}.
The IR symmetry we focus is:
\bea
\big(\frac{\SU(2) \times \Z_{8,\tA}}{\Z_2^F} \big) \times  \cancel{\U(1)^e_{[1]}}  \times  \cancel{\U(1)^m_{[1]}}
\eea
The \cancel{(\dots)} means a spontaneous symmetry breaking of $\dots$, thus for
1-form symmetry breaking here, it leads to a deconfinement of U(1) gauge theory.
In summary, the scenario \ref{s4} proposed that:
\bea
\text{``chiral symmetry preserved, and deconfinement.''}\quad 
\eea

\item \label{s5}
Note that there is another scenario from Ref.~\cite{Anber2018tcj1805.12290} proposing only
a free massless Dirac fermion at IR (equivalently, two massless Weyl fermions, or two massless Majorana fermions),
and two vacua (two degenerate ground states) due to $\Z_{8,\tA} \to \Z_{4,\tA}$, without any 1-form symmetry.
This scenario is certainly incomplete due to the lack of matching the higher 't Hooft anomalies of 1-form symmetry.
As Ref.~\cite{Anber2018tcj1805.12290} also notices later, 
the more complete scenario is adding a TQFT sector, following the Scenario \ref{s2}.

\end{enumerate}

\begin{widetext}

\onecolumngrid

\begin{table}[h!]
\footnotesize
    \hspace{-6mm}
    \begin{tabular}{| c | c | c | c | c |c|}
    \hline
   Scenario & $\begin{array}{c} \text{Internal global} \\  \text{symmetry } \mathbb{G}  \end{array}$  & 
   $\begin{array}{c}\text{Chiral}\\ \text{Symmetry}\end{array}$  &  
   $\begin{array}{c} \text{ 1-form $\Z_{2,[1]}^e$ Symmetry;} \\  \text{De-/Confinement} \end{array}$ 
   & $\begin{array}{c} \text{Anomaly}\\ \text{matched} \\ \text{with UV} \end{array}$  
     &  $\begin{array}{c} \text{Plausible} \\  \text{Candidates}  \end{array}$ 
   \\ \hline\hline
\ref{s1}. Ref.~\cite{2018arXiv180609592C} 
&
$\tO(2) \times  \U(1)_{[1]}$ &
$\begin{array}{c}\text{SSB}
\end{array}$
 & 
Enhanced and preserved; Confined.
& Yes & 
$\begin{array}{c} \text{Yes}\\
\text{(favored by energetic?)}  \end{array}$
\\
 \hline   
\ref{s2}. Ref.~\cite{2018arXiv180609592C} 
&
$\big(\frac{\SU(2) \times \Z_{4,\tA}}{\Z_2^F} \big) \times  \Z_{2,[1]}^e \times \dots
$ &  $\begin{array}{c}\text{SSB}\\
\Z_{8,\tA} \to \Z_{4,\tA}
\end{array}$ &
$\begin{array}{c}\text{Preserved; Confined.}\\ 
\text{But + new deconfined $\Z_2$-TQFT}\\
\text{with emergent new $\Z_{2,[1]}$ SSB.}
\end{array}$
& Yes & Yes\\
 \hline   
\ref{s3}. Ref.~\cite{Bi:2018xvr} 
&
$\big(\frac{\SU(2) \times \Z_{8,\tA}}{\Z_2^F} \big) \times  \Z_{2,[1]}^e$ 
& Preserved &
Preserved; Confined.
& $\begin{array}{c} \text{Obstruction} 
 \\  \text{of symmetric} \\
\text{TQFT}  \end{array}$
& $\begin{array}{c} \text{Obstruction.} 
 \\  \text{Not compatible w/} \\
\text{symmetry-extension  \cite{Wang2017loc170506728} }   \end{array}$
\\
\hline
\ref{s4}. Ref.~\cite{2018arXiv180609592C} 
&
$\big(\frac{\SU(2) \times \Z_{8,\tA}}{\Z_2^F} \big) \times  \cancel{\U(1)^e_{[1]}}  \times  \cancel{\U(1)^m_{[1]}}$ & Preserved &
Enhanced but SSB; Deconfined.
& Yes & Yes \\
 \hline   
    \end{tabular}
\caption{
The Scenario \ref{s1},  \ref{s2},   \ref{s3}, and  \ref{s4} are from the list summarized in \Sec{sec:fate-adjoint}. 
The ``SSB'' stands for ``spontaneous symmetry breaking.''
The \cancel{(\dots)} means that symmetry $(\dots)$ leads to SSB.
We find an obstruction for Scenario \ref{s3} based on the higher symmetry-extension construction of  Ref.~\cite{Wang2017loc170506728}.
We should note that, educated by Ref.~\cite{2018arXiv180609592C} and summarized in \Fig{Fig-3:SW}, 
the Scenario \ref{s1} is consistent with the supersymmetry (SUSY) breaking of
$\CN=2$ SYM from the magnetic monopole point and dyon point (as 2 copies of $\CP^1$ model).
The Scenario \ref{s2} is consistent with the SUSY breaking of
$\CN=2$ SYM from the generic point from $u \neq 0$, and $u \neq \pm \Lambda^2$.
The Scenario \ref{s4} is consistent with the SUSY breaking of
$\CN=2$ SYM from the $u = 0$ with a self-dual coupling {$\tau_{}$}.
  }
\label{table:adjoint}
\end{table}

\twocolumngrid

\end{widetext}

\subsection{
 Deconfined Quantum Criticality, Quantum Spin/Fermionic Liquids 
in 3+1 Dimensions
and
More Comments}
 
In this work, we obtain a higher-symmetry extension generalization of Ref.~\cite{Wang2017loc170506728}'s method
to construct symmetric anomalous TQFT saturating higher 't Hooft anomalies.
We have obtained a symmetric anomalous TQFT,
valid for Scenario \ref{s2} from  Cordova-Dumitrescu (Ref.~\cite{2018arXiv180609592C}), 
see 
\eq{eq:k2TypeI-II-2}
and
\eq{eq:TQFT-all}.
However, we are unable to obtain a symmetric anomalous TQFT
proposed by 
 Scenario \ref{s3} motivated by Bi-Senthil (Ref.~\cite{Bi:2018xvr}) based on a symmetry-extension construction.
 
 It is worthwhile to digest the exotic and interesting physics of Scenario \ref{s3} better.
The Scenario \ref{s3} is motivated by the
deconfined quantum criticality in 3+1 dimensions. It is proposed that a
critical theory can be realized as a phase transition between two conventional Landau-Ginzburg symmetry-breaking orders \cite{dQCP2004},
or a phase transition between two different SPT orders (see \cite{Bi:2018xvr} and References therein).
The adjoint QCD$_4$ is a UV description (UV side of \eq{eq:dual}) of the phase transition,
while the IR description is currently unclear (IR side of \eq{eq:dual}).

The novelty of Scenario \ref{s3} is that the gapless sector 
is a free CFT as two free Weyl fermions (a single free Dirac fermion). 
So the hope is that the possible UV-IR duality \eq{eq:dual} in 3+1D is between a strongly coupled and interacting UV gauge theory
and a free non-interacting massless IR theory, up to a gapped fully-symmetric TQFT sector to saturate the higher 't Hooft anomalies.

Our present work shows an obstruction for Scenario \ref{s3} from a symmetry-extension construction alone.
The implications of our finding are follows:
\noindent
\begin{enumerate}[leftmargin=*, label=\textcolor{blue}{\Roman*.}, ref=\Roman*]
\item
We should remind the readers that the symmetry-extension construction is fairly general enough to saturate a large class of
higher 't Hooft anomalies of bosonic systems. Although the adjoint QCD$_4$ is a fermionic system 
(the UV completion requires fermionic degrees of freedom, where there are gauge-invariant fermionic operators),
the Type I and II anomalies, \eq{eq:type1} and \eq{eq:type2}, are bosonic anomalies in nature.

\item Despite the fact that fully-symmetric TQFT under Scenario \ref{s3} cannot be obtained via our symmetry-extension construction,
we may still be able to use the symmetry-extension construction to derive other symmetric anomalous TQFTs, suitable to propose
new candidate phases of other \emph{deconfined quantum criticality} (dQCP), in 3+1 and other dimensions.
\end{enumerate}


We should also notice that the recent numerical attempts \cite{DelDebbio2009fd-0907.3896, Athenodorou2014eua1412.5994} suggest that
the adjoint QCD$_4$ with SU(2) gauge group and $N_f$ number of adjoint Weyl fermions may have IR dynamics as follows:
 \noindent
\begin{itemize}[leftmargin=*] 
\item  At $N_f=2$, (as 1 adjoint Dirac fermion), according to \cite{Athenodorou2014eua1412.5994},
the IR theory may be very close to the onset of the conformal window, instead of the conventional confining behavior.
In addition, the anomalous dimension of the fermionic condensate is reported to be close to 1.
The numerical data
seems to suggest the IR theory can be an interacting CFT (more exotic), instead of a free CFT (all the proposed scenarios so far, discussed in Table \ref{table:adjoint}).

\item  At $N_f=4$, (as 2 adjoint Dirac fermion), Ref.~\cite{DelDebbio2009fd-0907.3896}
discusses the candidate IR theory. 
Ref.~\cite{DelDebbio2009fd-0907.3896} points out the theory is gapless (or massless),
while future endeavor is required to distinguish whether it shows the confinement or the conformal behavior.
\end{itemize}
To unambiguously determine the IR dynamics, apart from the given numerical inputs \cite{DelDebbio2009fd-0907.3896, Athenodorou2014eua1412.5994},
we note that further lattice studies are still necessary.\\

In addition, the adjoint QCD$_4$ system has the $\Z_2^F$ fermionic parity symmetry
un-gauged; thus $\Z_2^F$ remained an honest global symmetry.
This system can be viewed as the emergent QFT of a
3+1 spacetime dimensional Quantum Fermionic Liquids ---
a fermionic analog of the familiar bosonic Quantum Spin Liquids (QSL).
In condensed matter, the ``spin'' in QSL implies the ``isospin'' which is bosonic in nature. While QSL does not require manifolds with spin structures to be realized,
the adjoint QCD$_4$ requires certain manifolds endorsed with analogs of spin structure
given in \eq{eq:Spin-sym} and \eq{eq:Pin-sym}. \\

Finally, we remark that many anomalies discussed in \Sec{sec:anom}, following \cite{2018arXiv180609592C},
are non-perturbative global anomalies instead of perturbative anomalies. The  non-perturbative anomalies have classifications from finite groups (e.g. $\Z_n$ classes), instead of
a $\Z$ classification.
Examples include the old and the new SU(2) anomalies \cite{Witten:1982fp, Wang:2018qoyWWW},
and also the recent higher 't Hooft anomalies of SU(N) YM gauge theory, see \cite{Gaiotto2017yupZoharTTT}  and \cite{Wan2018zql1812.11968}, and References therein.
For these non-perturbative global anomalies, we can saturate certain 't Hooft anomalies 
of ordinary or higher global symmetries by symmetry-preserving TQFTs or so-called the long-ranged entangled topological order sectors, via our 
higher symmetry-extension approach, see a companion work along this direction \cite{Wan2018zql1812.11968}.

\section{Acknowledgments}

%
The authors are listed in the alphabetical order by the standard convention.
JW thanks the participants of Developments in Quantum Field Theory and Condensed Matter Physics (November 5-7, 2018) 
at Simons Center for Geometry and Physics at SUNY Stony Brook University
for giving valuable feedback where this work is publicly reported \cite{SCGP}. 
We thank Edward Witten for helpful comments and sharing his proof from another perspective.
We are grateful to Chang-Tse Hsieh for Email correspondences on Ref.~\cite{Gilkey1996}, and informing his related result in 
Ref.~\cite{1808.02881}, such that we make a correction in our Table \ref{table4} and \ref{table5} after the Phys.~Rev.~D publication \cite{Wan2018djl1812.11955}.
JW especially thanks Clay Cordova, also thanks Kantaro Ohmori, Nathan Seiberg, and Shu-Heng Shao for illuminating conversations. 
Part of this short result emerges from an unpublished work in \cite{SWW}.
ZW  acknowledges support from NSFC grants 11431010 and 11571329. 
JW  acknowledges the Corning Glass Works Foundation Fellowship and NSF Grant PHY-1606531. 
This work is also supported by NSF Grant DMS-1607871 ``Analysis, Geometry and Mathematical
Physics'' and Center for Mathematical Sciences and Applications at Harvard University.

\appendix

\section{Cobordism Theory and Higher Symmetry-Extension: Construction of 
Symmetric TQFTs}

\label{sec:cobordism-TQFT}

By the result of the page 251 of
Ref.~\cite{hatcher},
the cohomology ring of the infinite Lens space $\B\Z_{2^n}=S^{\infty}/\Z_{2^n}$ with coefficients $\Z_{2^n}$ is the polynomial ring generated by $a$ and $b$ over $\Z_{2^n}$ quotient by the relation 
$a^2=2^{n-1}b$:
\be
\H^*(\B \Z_{2^n},\Z_{2^n})=\Z_{2^n}[a,b]/(a^2=2^{n-1}b) \;\;\;\text{for }n\ge2,
\ee
where $a\in\H^1(\B \Z_{2^n},\Z_{2^n})$, $b\in\H^2(\B \Z_{2^n},\Z_{2^n})$.
\bea
\H^*(\B \Z_{2^n},\Z_2)=\Lambda_{\Z_2}(\tilde{a})\otimes\Z_2[\tilde{b}] \;\;\;\text{for }n\ge2,
\eea
where $\tilde{a}=a\mod2$, $\tilde{b}=b\mod2$, there is a $(2,2^n)$-Bockstein $\beta_{(2,2^n)}$ with $\beta_{(2,2^n)}(a)=\tilde{b}$.
{Here $\H^*$ is the cohomology ring, $\Lambda_{\Z_2}$ denotes the exterior algebra over $\Z_2$, $\otimes$ is the tensor product. The $(2,2^n)$-Bockstein homomorphism $\beta_{(2,2^n)}:\H^*(-,\Z_{2^n})\to\H^{*+1}(-,\Z_2)$ is associated to the extension $\Z_2\to\Z_{2^{n+1}}\to\Z_{2^n}$.}

{Notice that the notation $a\in\H^1(\B\Z_{2^n},\Z_{2^n})$ and $b\in\H^2(\B\Z_{2^n},\Z_{2^n})$ will be abused later, since we will encounter the cases $n=2$ and $n=3$. We will use the uniform notation and explain wherever they appear.}

\subsection{Pullback trivialization of $A \mathcal{P}_2(B_2)$ in $\Omega_d^{\Spin\times_{\Z_2}(\SU(2)\times_{\Z_2}\Z_8)}(\B ^2\Z_2)$}\label{OmegaSpintimesZ2SU2timesZ2Z8B2Z2}

Follow the mathematical conventional notation, we will also denote
the 5d topological term 
\bea
A\cup\mathcal{P}_2(B_2) \text{ as } 
a\cup\mathcal{P}_2(x_2)
\eea
in \App{sec:cobordism-TQFT} and after. The $a$ here is a background probed field, which should not be confused with the
SU(2) dynamical gauge field.

\subsubsection{Computation}

$\Spin\times_{\Z_2}(\SU(2)\times_{\Z_2}\Z_8) \equiv (\Spin\times(\SU(2)\times\Z_8)/\Z_2)/\Z_2$ where the quotient is with respect to the diagonal center $\Z_2$ subgroup.

Since the computation involves no odd torsion, we can use the Adams spectral sequence
\bea
&&E_2^{s,t}=\Ext_{\A_2}^{s,t}(\H^*(MT(\Spin\times_{\Z_2}(\SU(2)\times_{\Z_2}\Z_8)),\Z_2)\otimes\notag\\
&&\H^*(\B ^2\Z_2,\Z_2),\Z_2)\Rightarrow\Omega_{t-s}^{\Spin\times_{\Z_2}(\SU(2)\times_{\Z_2}\Z_8)}(\B ^2\Z_2).
\eea
{
Here $\Ext$ is the Ext functor, $\A_2$ is the mod 2 Steenrod algebra, more precisely, $\Ext_{\A_2}^{s,t}$ is the internal degree $t$ part of the $s$-th derived functor of $\Hom_{\A_2}^*$. $MT(\Spin\times_{\Z_2}(\SU(2)\times_{\Z_2}\Z_8))$ is the Madsen-Tillmann spectrum of the group $\Spin\times_{\Z_2}(\SU(2)\times_{\Z_2}\Z_8)$, the bordism group $\Omega_d^{\Spin\times_{\Z_2}(\SU(2)\times_{\Z_2}\Z_8)}(\B^2\Z_2)=\pi_d(MT(\Spin\times_{\Z_2}(\SU(2)\times_{\Z_2}\Z_8))\wedge (\B^2\Z_2)_+)$ is the stable homotopy group of the spectrum $MT(\Spin\times_{\Z_2}(\SU(2)\times_{\Z_2}\Z_8))\wedge (\B^2\Z_2)_+$, here
$\wedge$ is the smash product, $X_+$ is the disjoint union of the space $X$ and a point. ``$\Rightarrow$'' means ``convergent to''. For more detail, see \cite{W2}.
}

Similarly as the discussion in \cite{2017arXiv170804264C,2017arXiv171111587GPW}, we know
\bea
&&MT(\Spin\times_{\Z_2}(\SU(2)\times_{\Z_2}\Z_8))\nn\\
&=&M\Spin\wedge\Sigma^{-3}M\SO(3)\wedge(\B \Z_4)^{2\xi}
\eea
where $2\xi$ is twice the sign representation, $(\B \Z_4)^{2\xi}$ is the Thom space $\text{Thom}(\B \Z_4;2\xi)$, $M\Spin$ is the Thom spectrum of the group $\Spin$, $M\SO(3)$ is the Thom spectrum of the group $\SO(3)$, $\Sigma$ is the suspension.

Note that $(\B \Z_4)^{2\xi}=\Sigma^{-2}M\Z_4$.

We have a homotopy pullback square

\bea\label{pullbacksquare}
\xymatrix{
\B (\Spin\times_{\Z_2}(\SU(2)\times_{\Z_2}\Z_8))  \ar[r] \ar[d]&\B\SO(3)\times\B \Z_4\ar[d]^{w_2'+\tilde{b}}\\
   \B \SO \ar[r]_{w_2}      &\B ^2\Z_2}\quad
\eea
where $\tilde b$ is the generator of $\H^2(\B\Z_4,\Z_2)$, $w_2=w_2(TM)$ is the Stiefel-Whitney class of the tangent bundle $TM$, $w_2'=w_2'(\SO(3))$ is the Stiefel-Whitney class of the universal $\SO(3)$ bundle.

Hence we have the constraint 
\bea
w_2(TM)=w_2'(\SO(3))+\tilde b
\eea

{Since $\H^*(M\Spin,\Z_2)=\A_2\otimes_{\A_2(1)}\{\Z_2\oplus M\}$
where $\A_2(1)$ is the sub-algebra of $\A_2$ generated by $\Sq^1$ and $\Sq^2$, and $M$ is a graded $\A_2(1)$-module with the degree $i$ homogeneous part $M_i = 0$ for $i < 8$.
}

For $t-s<8$, we can identify the $E_2$ page with
\bea
&&\Ext_{\A_2(1)}^{s,t}(\H^{*+3}(M\SO(3),\Z_2)\otimes\H^{*+2}(M\Z_4,\Z_2)\nn\\
&&\otimes\H^*(\B ^2\Z_2,\Z_2),\Z_2).
\eea


$\H^{*+3}(M\SO(3),\Z_2)=\Z_2[w_2',w_3']U$ where $U$ is the Thom class and $w_i'$ is the Stiefel-Whitney class of the universal $\SO(3)$ bundle.

\begin{figure}[!h]
\begin{center}
\includegraphics{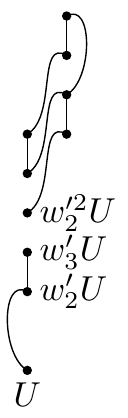}
\end{center}
\caption{The $\A_2(1)$-module structure of $\H^{*+3}(M\SO(3),\Z_2)$. 
{Each dot indicates a $\Z_2$, the short straight line indicates a $\Sq^1$, the curved line indicates a $\Sq^2$.}}
\label{fig:H3MSO3Z2}
\end{figure}

The $\A_2(1)$-module structure of $\H^{*+3}(M\SO(3),\Z_2)$ is shown in Figure \ref{fig:H3MSO3Z2}.

$\H^{*+2}(M\Z_4,\Z_2)=(\Z_2[\tilde b]\otimes\Lambda_{\Z_2}(\tilde a))U$ where $U$ is the Thom class, $\tilde{a}$ is the generator of $\H^1(\B \Z_4,\Z_2)$, $\tilde{b}$ is the generator of $\H^2(\B \Z_4,\Z_2)$.

\begin{figure}[!h]
\begin{center}
\includegraphics{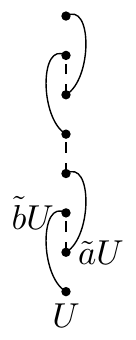}
\end{center}
\caption{The $\A_2(1)$-module structure of $\H^{*+2}(M\Z_4,\Z_2)$. The dashed lines indicate a $(2,4)$-Bockstein. 
{Each dot indicates a $\Z_2$, the short straight line indicates a $\Sq^1$, the curved line indicates a $\Sq^2$.}}
\label{fig:H2MZ4Z2}
\end{figure}

The $\A_2(1)$-module structure of $\H^{*+2}(M\Z_4,\Z_2)$ is shown in Figure \ref{fig:H2MZ4Z2}.

{$\H^*(\B^2\Z_2,\Z_2)=\Z_2[x_2,x_3,x_5,\dots]$ where $x_2$ is the generator of $\H^2(\B^2\Z_2,\Z_2)$, $x_3=\Sq^1x_2$, $x_5=\Sq^2x_3$, etc.}

\begin{figure}[!h]
\begin{center}
\includegraphics{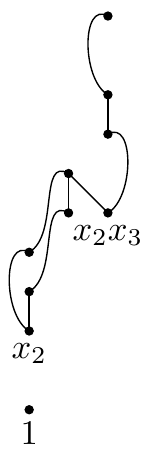}
\end{center}
\caption{The $\A_2(1)$-module structure of $\H^*(\B ^2\Z_2,\Z_2)$. 
{Each dot indicates a $\Z_2$, the short straight line indicates a $\Sq^1$, the curved line indicates a $\Sq^2$.}}
\label{fig:HBZ2Z2}
\end{figure}

The $\A_2(1)$-module structure of $\H^*(\B ^2\Z_2,\Z_2)$ is shown in Figure \ref{fig:HBZ2Z2}.

\begin{figure}[!h]
\begin{center}
\includegraphics{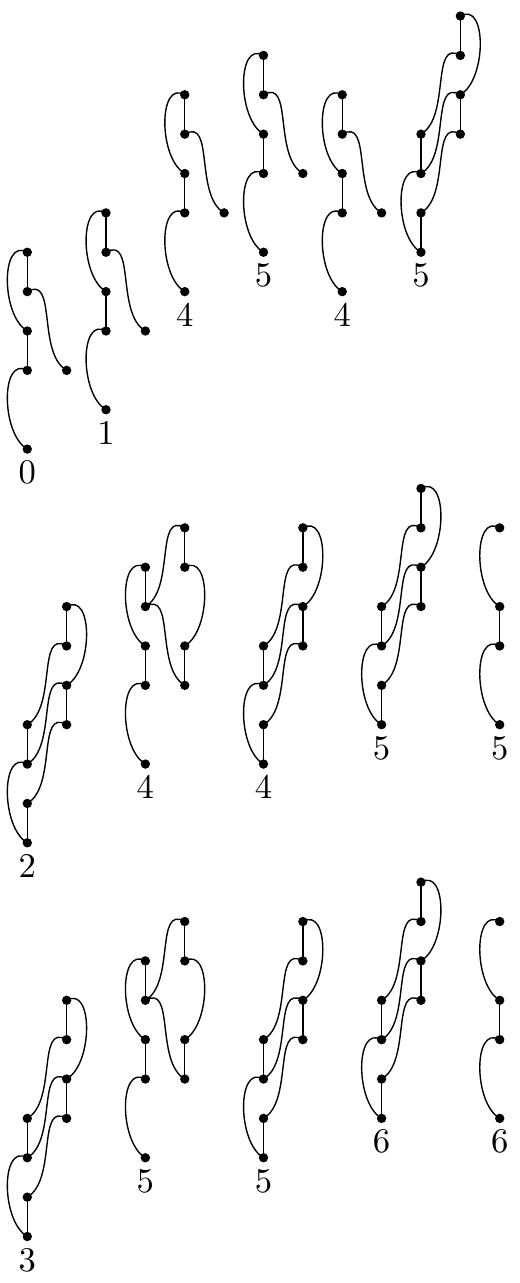}
\end{center}
\caption{The $\A_2(1)$-module structure of $\H^{*+3}(M\SO(3),\Z_2)\otimes\H^{*+2}(M\Z_4,\Z_2)\otimes\H^*(\B ^2\Z_2,\Z_2)$. 
{Each dot indicates a $\Z_2$, the short straight line indicates a $\Sq^1$, the curved line indicates a $\Sq^2$. Each label indicates its degree.}}
\label{fig:H3MSO3Z2otimesH2MZ4Z2otimesHBZ2Z2}
\end{figure}

The $\A_2(1)$-module structure of $\H^{*+3}(M\SO(3),\Z_2)\otimes\H^{*+2}(M\Z_4,\Z_2)\otimes\H^*(\B ^2\Z_2,\Z_2)$ is shown in Figure \ref{fig:H3MSO3Z2otimesH2MZ4Z2otimesHBZ2Z2}.

There is a differential $d_2$ corresponds to the $(2,4)$-Bockstein \cite{may1981bockstein} as indicated in Figure \ref{fig:H2MZ4Z2}.
There is also a differential $d_2$ maps $x_2x_3+x_5$ to $x_2^2h_0^2$ since $\beta_{(2,4)}(\mathcal{P}_2(x_2))=x_2x_3+x_5$ 
{\cite{W2}}.
Since $\beta_{(2,4)}(a\mathcal{P}_2(x_2))=\tilde{b}x_2^2+\tilde{a}(x_2x_3+x_5)$, there is a differential $d_2$ maps $\tilde{b}x_2^2+\tilde{a}(x_2x_3+x_5)$ to $\tilde{a}x_2^2h_0^2$.

\begin{figure}[!h]
\begin{center}
\includegraphics{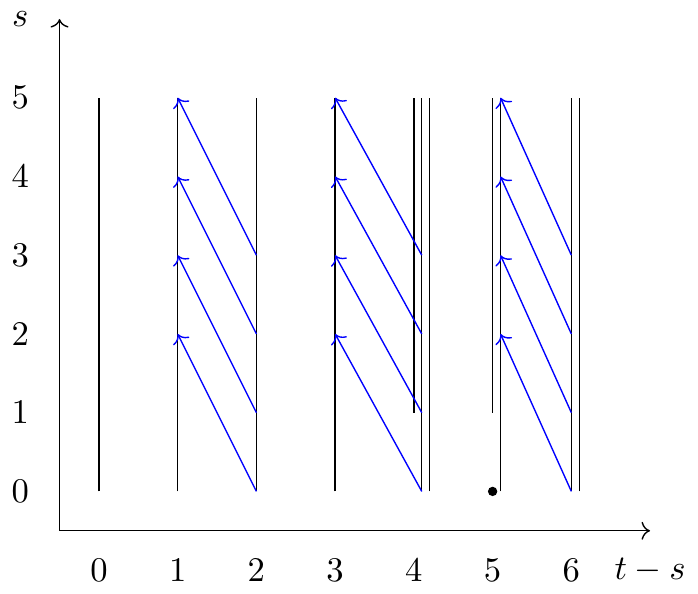}
\end{center}
\caption{$\Omega_*^{\Spin\times_{\Z_2}(\SU(2)\times_{\Z_2}\Z_8)}$. 
{The arrows indicate differentials.}}
\label{fig:OmegaSpintimesZ2SU2timesZ2Z8}
\end{figure}

\begin{figure}[!h]
\begin{center}
\includegraphics{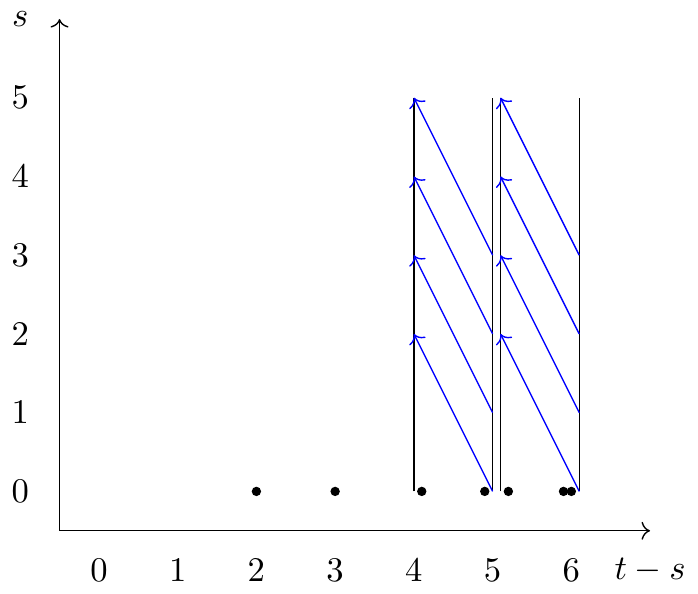}
\end{center}
\caption{$(\Omega_*^{\Spin\times_{\Z_2}(\SU(2)\times_{\Z_2}\Z_8)}(\B ^2\Z_2))/(\Omega_*^{\Spin\times_{\Z_2}(\SU(2)\times_{\Z_2}\Z_8)})$. 
{The arrows indicate differentials.}}
\label{fig:OmegaSpintimesZ2SU2timesZ2Z8B2Z2}
\end{figure}

Note that the $\A_2(1)$-module structure of $\H^{*+3}(M\SO(3),\Z_2)\otimes\H^{*+2}(M\Z_4,\Z_2)$ is contained in that of $\H^{*+3}(M\SO(3),\Z_2)\otimes\H^{*+2}(M\Z_4,\Z_2)\otimes\H^*(\B ^2\Z_2,\Z_2)$, we draw the $E_2$ page for it individually in Figure \ref{fig:OmegaSpintimesZ2SU2timesZ2Z8}.
The rest part is shown in Figure \ref{fig:OmegaSpintimesZ2SU2timesZ2Z8B2Z2}.

See Table \ref{table2} for the bordism group data.

\begin{table}[!h]
\centering
\begin{tabular}{c c c}
\hline
$i$ & $\Omega_i^{\Spin\times_{\Z_2}(\SU(2)\times_{\Z_2}\Z_8)}$ & cobordism invariants\\
\hline
0& $\Z$ \\
1& $\Z_4$ & $a$\\
2& $0$\\
3 & $\Z_4$ & $ab$\\
4 & $\Z^2$\\
5 & $\Z\times\Z_2\times\Z_4$ & $w_2'w_3',\ ab^2$\\
\hline
\end{tabular}
\caption{Bordism group $\Omega_i^{\Spin\times_{\Z_2}(\SU(2)\times_{\Z_2}\Z_8)}$ in dimensions $i$. Here $a$ is the generator of $\H^1(\B\Z_4,\Z_4)$, $b$ is the generator of $\H^2(\B\Z_4,\Z_4)$. $w_i'$ is the Stiefel-Whitney class of the universal $\SO(3)$ bundle.}
\label{table2}
\end{table}

See Table \ref{table3} for the bordism group data.

\begin{table}[!h]
\centering
\hspace{-2.8em}
\begin{tabular}{c c c}
\hline
$i$ & $\Omega_i^{\Spin\times_{\Z_2}(\SU(2)\times_{\Z_2}\Z_8)}(\B ^2\Z_2)$ & cobordism invariants\\
\hline
0& $\Z$\\
1& $\Z_4$ & $a$\\
2& $\Z_2$ & $x_2$\\
3 & $\Z_2\times\Z_4$ & $\tilde a x_2,\ ab$\\
4 & $\Z^2\times\Z_2\times\Z_4$ & $\tilde b x_2,\ \mathcal{P}_2(x_2)$\\
5 & $\Z\times\Z_2^3\times\Z_4^2$ & $w_2'w_3',\ \tilde a \tilde b x_2, \ w_2x_3,\ ab^2,\ a\mathcal{P}_2(x_2)$\\
\hline
\end{tabular}
\caption{Bordism group $\Omega_i^{\Spin\times_{\Z_2}(\SU(2)\times_{\Z_2}\Z_8)}(\B ^2\Z_2)$ in dimensions $i$. Here $a$ is the generator of $\H^1(\B\Z_4,\Z_4)$, $b$ is the generator of $\H^2(\B\Z_4,\Z_4)$. $\tilde a=a\mod2$, $\tilde b=b\mod2$.
$w_2=w_2(TM)$ is the Stiefel-Whitney class of the tangent bundle. Note that $w_2x_3=w_3x_2$ (see \cite{W2}). $w_i'$ is the Stiefel-Whitney class of the universal $\SO(3)$ bundle.}
\label{table3}
\end{table}


\subsubsection{Manifold generator}

Now we determine the manifold generator of the $\Z_4$-valued invariant $a\cup\mathcal{P}_2(x_2)$.

\bea
&&\Omega_5^{\Spin\times_{\Z_2}(\SU(2)\times_{\Z_2}\Z_8)}(\B ^2\Z_2)\notag\\
&=&\{5\text{-manifolds }M\text{ with maps }\nn\\
&&f:M\to \B (\Spin\times_{\Z_2}(\SU(2)\times_{\Z_2}\Z_8))\notag\\
&&\text{ and }g:M\to \B ^2\Z_2\}/\text{bordism}
\eea
{
Here bordism is an equivalence relation. $(M,f,g)$ and $(M',f',g')$ are bordant if there exists a 6-manifold $\CM$ and maps $F:\CM\to \B (\Spin\times_{\Z_2}(\SU(2)\times_{\Z_2}\Z_8))$, $G:\CM\to \B ^2\Z_2$
such that the boundary of $\CM$ is the disjoint union of $M$ and $M'$ and the induced $\Spin\times_{\Z_2}(\SU(2)\times_{\Z_2}\Z_8)$ structures on $M$ and $M'$ from that determined by $F$ on $\CM$ coincide with those determined by $f$ and $f'$ respectively, and $G|_M=g$, $G|_{M'}=g'$. 
}

We have the homotopy pullback square \eqref{pullbacksquare}.


In order to give a map $f:M\to \B (\Spin\times_{\Z_2}(\SU(2)\times_{\Z_2}\Z_8))$, we need only give maps $f_1:M\to \B \SO$, $f_2:M\to \B\SO(3)$ and $f_3:M\to \B \Z_4$ with $f_1^*(w_2)=f_2^*(w_2')+f_3^*(\tilde{b})$.

The bordism invariant $a\cup\mathcal{P}_2(x_2)$ is actually $f_3^*(a)\cup\mathcal{P}_2(g^*(x_2))=f_3\cup\mathcal{P}_2(g)$. 

Now let $M$ be the Lens space $S^5/\Z_4$, $M$ is orientable but not spin.

Take $f_1=TM$ (since $M$ is orientable, the tangent bundle $TM$ determines a map $M\to\B\SO$), $f_2=0$, $f_3$ be the generator of $\H^1(M,\Z_4)$.

By the cell structure of the Lens space, $f_3$ induces a chain map between the cellular chain complexes of $M$ and $\B \Z_4$, we draw the chain map below degree 2:
\bea
\xymatrix{
\Z \ar[r]^4 \ar[d]^1 & \Z\ar[r]^0 \ar[d]^1 & \Z\ar[d]\\
\Z \ar[r]^4      &\Z\ar[r]^0 & \Z.}
\eea

So $f_3^*(\tilde{b})$ is nonzero, since $f_1^*(w_2)$ is also nonzero, the cohomology group $\H^2(M,\Z_2)$ is $\Z_2$,
we have a commutative diagram
\bea
\xymatrix{
M \ar[r]^{f_3} \ar[d]_{f_1}&\B \Z_4\ar[d]^{\tilde{b}}\\
   \B \SO \ar[r]_{w_2}      &\B ^2\Z_2.}
\eea
So we get a map $f:M\to \B (\Spin\times_{\Z_2}(\SU(2)\times_{\Z_2}\Z_8))$.

Take $g=w_2(TM)$, and
\bea
\int_{M}f_3\cup \mathcal{P}_2(g)=1\mod4.
\eea

The partition function is
\bea
Z(M)=\text{i}^{\int_{M} f_3   \cup \mathcal{P}_2(g)}=\text{i}.
\eea

So $(M,f,g)$ is the manifold generator of the $\Z_4$-valued invariant $f_3\cup\mathcal{P}_2(g)$.

\subsection{Pullback trivialization}

\label{subsec:pullback}

Consider the pullback of $\B (\Spin\times_{\Z_2}(\SU(2)\times_{\Z_2}\Z_8))$ to $\B \Spin\times \B (\SU(2)\times_{\Z_2}\Z_8)$:
\bea
&&\B \Z_2\to \B \Spin\times \B (\SU(2)\times_{\Z_2}\Z_8)\to \nn\\
&& \B (\Spin\times_{\Z_2}(\SU(2)\times_{\Z_2}\Z_8)).
\eea

Since $w_2=0$ in Spin, so $w_2x_3=w_3x_2$ is trivialized.

%
%
%
Furthermore, consider the pullback of $\B \Spin\times \B (\SU(2)\times_{\Z_2}\Z_8)$ to $\B\Spin\times\B\SU(2)\times\B\Z_8$:
\bea
&&\B \Z_2\to \B\Spin\times\B\SU(2)\times\B\Z_8\to \nn\\
&& \B \Spin\times \B (\SU(2)\times_{\Z_2}\Z_8).
\eea

To simplify the computation, we only compute $\Omega_5^{\Spin}(\B \Z_8\times \B ^2\Z_2)$ which is a subgroup of $\Omega_5^{\Spin}(\B\SU(2)\times\B \Z_8\times \B ^2\Z_2)$.

{
Note that 
$\mathcal{P}_2(x_2)=x_2^2=\Sq^2(x_2)=(w_2(TM)+w_1(TM)^2)x_2=0\mod2$
on Spin manifolds where we have used the Wu formula, so $\mathcal{P}_2(x_2)$ can be divided by 2.
}

\subsubsection{Computation}

We have the Adams spectral sequence
\bea
&&E_2^{s,t}=\Ext_{\A_2}^{s,t}(\H^*(M\Spin,\Z_2)\otimes\H^*(\B \Z_8,\Z_2)\otimes\notag\\
&&\H^*(\B ^2\Z_2,\Z_2),\Z_2)\Rightarrow\Omega_{t-s}^{\Spin}(\B \Z_8\times \B ^2\Z_2).
\eea
For $t-s<8$,
\bea
&&\Ext_{\A_2(1)}^{s,t}(\H^*(\B \Z_8,\Z_2)\otimes\H^*(\B ^2\Z_2,\Z_2),\Z_2)\notag\\
&&\Rightarrow\Omega_{t-s}^{\Spin}(\B \Z_8\times \B ^2\Z_2).
\eea

\begin{figure}[!h]
\begin{center}
\includegraphics{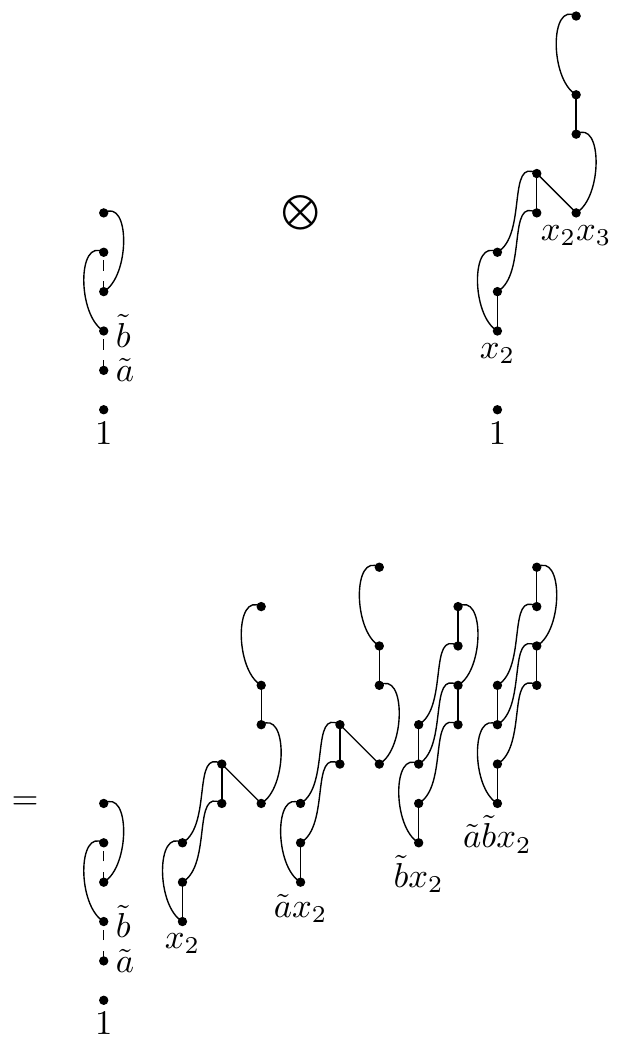}
\end{center}
\caption{The $\A_2(1)$-module structure of $\H^*(\B \Z_8,\Z_2)\otimes\H^*(\B ^2\Z_2,\Z_2)$. The dashed lines indicate an $(2,8)$-Bockstein. 
{Each dot indicates a $\Z_2$, the short straight line indicates a $\Sq^1$, the curved line indicates a $\Sq^2$.}}
\label{fig:HBZ8Z2otimesHB2Z2Z2}
\end{figure}

The $\A_2(1)$-module structure of $\H^*(\B \Z_8,\Z_2)\otimes\H^*(\B ^2\Z_2,\Z_2)$ is shown in Figure \ref{fig:HBZ8Z2otimesHB2Z2Z2}.

\begin{figure}[!h]
\begin{center}
\includegraphics{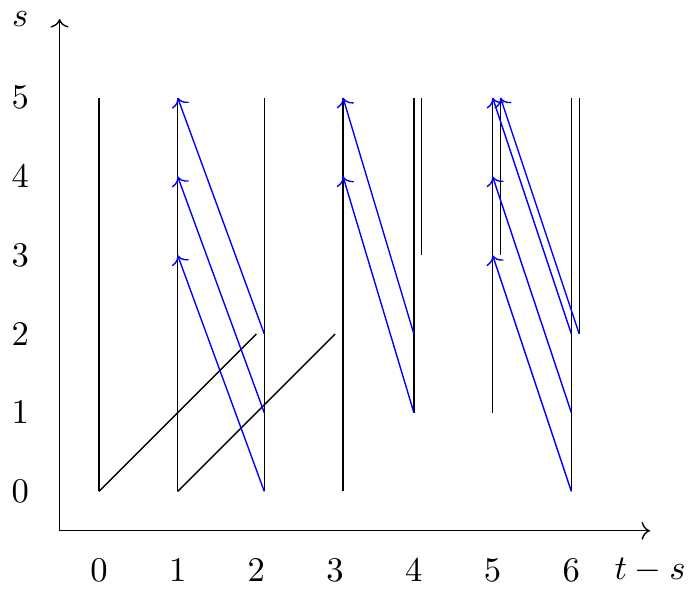}
\end{center}
\caption{$\Omega_*^{\Spin}(\B \Z_8)$. 
{The arrows indicate differentials.} Note that there is a nontrivial extension at $t-s=5$, the result is $\Omega_5^{\Spin}(\B\Z_8)=\Z_8\times\Z_2$, which has already been calculated in \cite{Gilkey1996} and \cite{1808.02881}. }
\label{fig:OmegaSpinBZ8}
\end{figure}

\begin{figure}[!h]
\begin{center}
\includegraphics{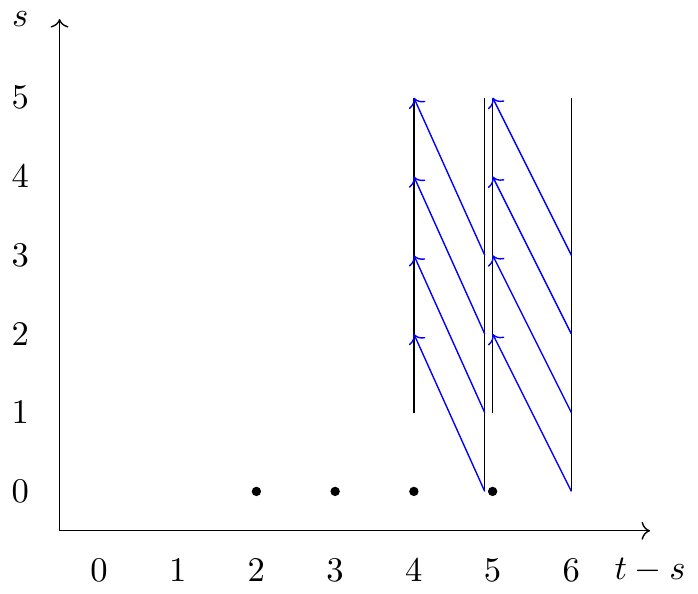}
\end{center}
\caption{$(\Omega_*^{\Spin}(\B \Z_8\times \B ^2\Z_2))/(\Omega_*^{\Spin}(\B \Z_8))$. 
{The arrows indicate differentials.}}
\label{fig:OmegaSpinBZ8timesB2Z2}
\end{figure}

Note that the $\A_2(1)$-module structure of $\H^*(\B \Z_8,\Z_2)$ is contained in that of $\H^*(\B \Z_8,\Z_2)\otimes\H^*(\B ^2\Z_2,\Z_2)$, we draw the $E_2$ page for it individually in Figure \ref{fig:OmegaSpinBZ8}.
The rest part is shown in Figure \ref{fig:OmegaSpinBZ8timesB2Z2}.

There is a differential $d_3$ corresponding to the $(2,8)$-Bockstein \cite{may1981bockstein} as indicated in Figure \ref{fig:HBZ8Z2otimesHB2Z2Z2} and a differential $d_2$ corresponding to the $(2,4)$-Bockstein 
$\beta_{(2,4)}(\mathcal{P}_2(x_2))=x_2x_3+x_5$.

See Table \ref{table4} for the bordism group data.

\begin{table}[!h]
\centering
\begin{tabular}{c c c}
\hline
$i$ & $\Omega_i^{\Spin}(\B \Z_8)$ & cobordism invariants\\
\hline
0& $\Z$\\
1& $\Z_2\times\Z_8$ & $\tilde\eta$, $a$\\
2& $\Z_2^2$ & $\tilde a\tilde\eta$, Arf\\
3 & $\Z_2\times\Z_{16}$ & $\tilde a$Arf, $\mathfrak{P}(a)$\\
4 & $\Z$\\
5 & $\Z_8\times\Z_2$\footnote{This bordism group and its cobordism invariants have already been calculated in \cite{Gilkey1996, 1808.02881}.
We are grateful to Chang-Tse Hsieh for informing us his result \cite{1808.02881}.} & 
\\
\hline
\end{tabular}
\caption{Bordism group $\Omega_i^{\Spin}(\B \Z_8)$ in dimensions $i$. Here $\tilde\eta$ is the 1d eta invariant, Arf is the Arf invariant, $\mathfrak{P}$ is the Postnikov square. $a$ is the generator of $\H^1(\B\Z_8,\Z_8)$, $b$ is the generator of $\H^2(\B\Z_8,\Z_8)$. $\tilde a=a\mod2$, $\tilde b=b\mod2$.
}
\label{table4}
\end{table}

See Table \ref{table5} for the bordism group data. 

\begin{table}[!h]
\centering
\begin{tabular}{c c c}
\hline
$i$ & $\Omega_i^{\Spin}(\B \Z_8\times \B ^2\Z_2)$ & cobordism invariants\\
\hline
0& $\Z$\\
1& $\Z_2\times\Z_8$ & $\tilde\eta$, $a$\\
2& $\Z_2^3$ & $\tilde a\tilde\eta$, Arf, $x_2$\\
3 & $\Z_2^2\times\Z_{16}$ & $\tilde a$Arf, $x_3$, $\mathfrak{P}(a)$\\
4 & $\Z\times\Z_2^2$ & $\tilde bx_2$, $\frac{\mathcal{P}_2(x_2)}{2}$\\
5 & $\Z_2^2\times\Z_8\times\Z_2$ & $\tilde a\tilde bx_2$, $\tilde a\frac{\mathcal{P}_2(x_2)}{2}$, \footnote{We omit the cobordism invariants of $\Omega_5^{\Spin}(\B\Z_8)=\Z_8\times\Z_2$ which have already been calculated in \cite{Gilkey1996} and \cite{1808.02881}.}\\
\hline
\end{tabular}
\caption{Bordism group $\Omega_i^{\Spin}(\B \Z_8\times \B ^2\Z_2)$ in dimensions $i$. Here $\tilde\eta$ is the 1d eta invariant, Arf is the Arf invariant, $\mathfrak{P}$ is the Postnikov square. $a$ is the generator of $\H^1(\B\Z_8,\Z_8)$, $b$ is the generator of $\H^2(\B\Z_8,\Z_8)$. $\tilde a=a\mod2$, $\tilde b=b\mod2$. }
\label{table5}
\end{table}

One $\Z_2$-valued bordism invariant of $\Omega_5^{\Spin}(\B \Z_8\times \B ^2\Z_2)$ is $\tilde{a}\cup \frac{\mathcal{P}_2(x_2)}{2}$.
Here $\tilde{a}$ is the generator of $\H^1(\B \Z_8,\Z_2)$, $x_2$ is the generator of $\H^2(\B ^2\Z_2,\Z_2)$.

\subsubsection{Further trivialization: first approach}

\label{subsec:1}

Define $\mathbb{G}$ to be a group which sits in a homotopy pullback square
%
\be
\xymatrix{
\B \mathbb{G}  \ar[rr] \ar[dd]&&\B ^2\Z_2\ar[dd]^{x_2}\\
&\B\SO(3)\times\B \Z_4\ar[rd]^{w_2'+\tilde{b}}&\\
   \B (\Spin\times_{\Z_2}(\SU(2)\times_{\Z_2}\Z_8)) \ar[ru]^{j_2}\ar[rd]_{j_1}&     &\B ^2\Z_2\\
   &\B \SO\ar[ru]_{w_2}&}
\ee
%
where $j_1^*(w_2)=j_2^*(w_2'+\tilde{b})$, $\tilde b$ is the generator of $\H^2(\B\Z_4,\Z_2)$, $w_2=w_2(TM)$ is the Stiefel-Whitney class of the tangent bundle $TM$, $w_2'=w_2'(\SO(3))$ is the Stiefel-Whitney class of the universal $\SO(3)$ bundle.

In general, if we have a homotopy pullback square
\bea
\xymatrix{P\ar[r]\ar[d]&Y\ar[d]\\
X\ar[r]&Z,}
\eea
then there is a fiber sequence
\bea
\Omega Z\to P\to X\times Y\to Z
\eea
where $\Omega Z$ is the loop space of $Z$. 
 
So there is a fiber sequence 
\bea
&&\B \Z_2\to \B \mathbb{G}\to \B (\Spin\times_{\Z_2}(\SU(2)\times_{\Z_2}\Z_8))\times \B ^2\Z_2\nn\\
&&\to \B ^2\Z_2,
\eea 
where the last map is $(u,v)\mapsto j_1^*(w_2)(u)-x_2(v)=j_2^*(w_2'+\tilde{b})(u)-x_2(v)$. 

Then we define $\mathbb{G}'$ to be a group which sits in a homotopy pullback square
\be
\xymatrix{
\B\Z_2\ar[r]&\B \mathbb{G}'\ar[r]\ar[d]&\B \Spin\times \B (\SU(2)\times_{\Z_2}\Z_8)\times \B ^2\Z_2\ar[d]\\
\B\Z_2\ar[r]&\B \mathbb{G}\ar[r] &\B (\Spin\times_{\Z_2}(\SU(2)\times_{\Z_2}\Z_8))\times \B ^2\Z_2.}
\ee

Since $w_2$ is identified with $x_2$ in $\B \mathbb{G}$, it is trivialized in $\B \mathbb{G}'$ 
because $x_2=w_2=0$ due to the spin structure,
so $a\cup \mathcal{P}_2(x_2)$ is clearly trivialized by being pulled back to $\Omega_5^{\mathbb{G}'}$. 

{Although our starting point was the symmetry-extension, this turns out to be a symmetry breaking case in disguise}.

\subsubsection{Further trivialization: second approach}

\label{subsec:2}

Define $\mathbb{G}$ to be a group which sits in a homotopy pullback square
\bea
\xymatrix{
\B \mathbb{G}\ar[d]\ar[rr]&&\B ^2\Z_2\ar[d]^{x_2}\\
\B (\Spin\times\SU(2)\times\Z_8)\ar[r]^-{j_3}&\B \Z_8\ar[r]^{\tilde{b}}&\B ^2\Z_2.}\quad
\eea

In general, if we have a homotopy pullback square
\bea
\xymatrix{P\ar[r]\ar[d]&Y\ar[d]\\
X\ar[r]&Z,}
\eea
then there is a fiber sequence
\bea
\Omega Z\to P\to X\times Y\to Z
\eea
where $\Omega Z$ is the loop space of $Z$. 

So there is a fiber sequence
\be
\B \Z_2\to \B \mathbb{G}\to \B (\Spin\times\SU(2)\times\Z_8)\times \B ^2\Z_2\to \B ^2\Z_2
\ee
where the last map is $(u,v)\mapsto j_3^*(\tilde{b})(u)-x_2(v)$.

Since $\mathcal{P}_2(x_2)=x_2\hcup{0}x_2+x_2\hcup{1}\delta x_2$, $\delta x_2=2\Sq^1x_2$, $x_2\hcup{0}x_2=x_2^2$, so $\frac{\mathcal{P}_2(x_2)}{2}=\frac{x_2^2}{2}+x_2\hcup{1}\Sq^1x_2$.

Since $x_2$ is identified with $\tilde{b}=\beta_{(2,8)}a$ in $\B \mathbb{G}$ where $a\in\H^1(\B\Z_8,\Z_8)$ and $\Sq^1\beta_{(2,8)}=0$ 
{\cite{W2}}, 
so $\tilde{a}\cup(x_2\hcup{1}\Sq^1x_2)$ is trivialized in $\Omega_5^{\mathbb{G}}$.

{Note that $\tilde{a}\cup\frac{x_2^2}{2}$ is still not trivialized.}

This is also a symmetry breaking case, 
{since $x_2$ is locked with $a$.
In physics, the locking between two probed background fields means that the global symmetry between two sectors are locked together, thus which results in 
global symmetry breaking.}

\subsubsection{Further trivialization: third approach}
\label{sec:third}

\label{subsec:3}

Consider the pullback of $\B ^2\Z_2$ to $\B ^2\Z_4$:
\bea
\B ^2\Z_2\to \B ^2\Z_4\to \B ^2\Z_2.
\eea

Since $\frac{\mathcal{P}_2(x_2)}{2}=\frac{x_2^2}{2}+x_2\hcup{1}\Sq^1x_2$, $x_2\in\H^2(\B ^2\Z_2,\Z_2)$ is pulled back to $\tilde{x}_2\in\H^2(\B ^2\Z_4,\Z_2)$ and the following diagram 
\bea
\xymatrix{\H^2(\B ^2\Z_4,\Z_2)\ar[r]^{\Sq^1}\ar[d]^{\cdot2}&\H^3(\B ^2\Z_4,\Z_2)\ar[d]^{\text{id}}\\
\H^2(\B ^2\Z_4,\Z_4)\ar[r]^{\beta_{(2,4)}}&\H^3(\B ^2\Z_4,\Z_2)}
\eea
is commutative by the naturality of Bockstein homomorphism, we have $\Sq^1\tilde{x}_2=0$, so $\tilde{a}\cup(x_2\hcup{1}\Sq^1x_2)$ is trivialized in $\Omega_5^{\Spin}(\B \Z_8\times \B ^2\Z_4)$.

{Note that $\tilde{a}\cup\frac{x_2^2}{2}$ is still not trivialized.}

\subsubsection{Summary 
}
The term $\tilde{a}\cup\frac{x_2^2}{2}$ cannot be trivialized.

Consider $M=S^1\times S^2\times S^2$, the partition function is
\bea
Z(M)=(-1)^{k\int_M\tilde{a}\cup\frac{x_2^2}{2}}=(-1)^{k\int_{S^2\times S^2}\frac{x_2^2}{2}}.
\eea
Since 
\bea
\H^n(S^2\times S^2,\Z)=\left\{\begin{array}{lll}\Z^2&n=2\\\Z&n=0,4\\0&n=1,3,n\ge4\end{array}\right.
\eea
where the two generators of $\H^2(S^2\times S^2,\Z)$ are $a,b$, the generator of $\H^4(S^2\times S^2,\Z)$ is $ab$.

No matter how to pullback, when $x_2=a+b\mod2$, $(-1)^{k\int_{S^2\times S^2}\frac{x_2^2}{2}}=(-1)^k$ can be nontrivial.

{This conclusion will be stated more formally and proved in the next appendix.}

{In this appendix, we compute the bordism group $\Omega_5^{\Spin\times_{\Z_2}(\SU(2)\times_{\Z_2}\Z_8)}(\B^2\Z_2)$ and find a bordism invariant $a\cup\mathcal{P}_2(x_2)$ of it. Then we find the manifold generator of $a\cup\mathcal{P}_2(x_2)$, and consider the pullback trivialization problem of $a\cup\mathcal{P}_2(x_2)$. 
We first compute the bordism group $\Omega_5^{\Spin}(\B\Z_8\times\B^2\Z_2)$, and find that $a\cup\mathcal{P}_2(x_2)$ becomes $\tilde a\cup\frac{\mathcal{P}_2(x_2)}{2}$ in $\Omega_5^{\Spin}(\B\Z_8\times\B^2\Z_2)$.
Moreover, we find that the summand 
$\tilde a \cup(x_2\hcup{1}\Sq^1x_2)$ of $\tilde a\cup\frac{\mathcal{P}_2(x_2)}{2}$ can be trivialized (note that $\mathcal{P}_2(x_2)=x_2^2+2x_2\hcup{1}\Sq^1x_2$), but $\tilde a \cup\frac{x_2^2}{2}$ cannot be trivialized.
We conclude that $a\cup\mathcal{P}_2(x_2)$ cannot be trivialized via extending the global symmetry by 0-form symmetry and 1-form symmetry.}

\section{Proof: a counterexample}

\label{sec:proof}

By direct computation, we find that $a\cup\mathcal{P}_2(x_2)$ is a bordism invariant of $\Omega_5^{\Spin\times_{\Z_2}(\SU(2)\times_{\Z_2}\Z_8)}(\B^2\Z_2)$.

We consider the trivialization problem: Can we trivialize the topological term $a\cup\mathcal{P}_2(x_2)$ via extending the global symmetry by 0-form $K_{[0]}$ symmetry and 1-form $K_{[1]}$ symmetry?

We can reformulate it mathematically: Can we find find finite abelian groups $K_{[0]}$ and $K_{[1]}$ such that 
\be
{\B K_{[0]} \ltimes \B^2 K_{[1]}}\to \B\mathbb{G}\stackrel{f}{\to} \B(\Spin\times_{\Z_2}(\SU(2)\times_{\Z_2}\Z_8))\times\B^2\Z_2
\ee
is a fibration and $(fg)^*(a\cup\mathcal{P}_2(x_2))=0$ for any 5-manifold $M$ and any map $g:M\to \B\mathbb{G}$?
 

There is a group homomorphism:
\bea
\Omega_5^{\mathbb{G}}&\stackrel{\phi}{\to}&\Omega_5^{\Spin\times_{\Z_2}(\SU(2)\times_{\Z_2}\Z_8)}(\B^2\Z_2)\notag\\
(M,g)&\mapsto&(M,fg)
\eea

So the trivialization problem is asking whether we can find $\mathbb{G}$ and $f$ such that $\phi^*(a\cup\mathcal{P}_2(x_2))=0$ for any $(M,g)\in\Omega_5^{\mathbb{G}}$.

By direct computation, we find that $a\cup\mathcal{P}_2(x_2)$ becomes $\tilde{a}\cup\frac{\mathcal{P}_2(x_2)}{2}$ in $\Omega_5^{\Spin}(\B\Z_8\times\B^2\Z_2)$.

Our main result is

Claim 1: We cannot find finite abelian groups $K_{[0]}$ and $K_{[1]}$ such that 
\be
{\B K_{[0]} \ltimes \B^2 K_{[1]}}\to \B\mathbb{G}\stackrel{f}{\to} \B(\Spin\times_{\Z_2}(\SU(2)\times_{\Z_2}\Z_8))\times\B^2\Z_2
\ee
is a fibration and $(fg)^*(a\cup\mathcal{P}_2(x_2))=0$ for any 5-manifold $M$ and any map $g:M\to \B\mathbb{G}$.

Claim 2: We cannot find finite abelian groups $K_{[0]}$ and $K_{[1]}$ such that 
\be
{\B K_{[0]} \ltimes \B^2 K_{[1]}}\to \B\mathbb{G}\stackrel{f}{\to} \B\Spin\times\B\SU(2)\times\B\Z_8\times\B^2\Z_2
\ee
is a fibration and $(fg)^*(\tilde{a}\cup\frac{\mathcal{P}_2(x_2)}{2})=0$ for any 5-manifold $M$ and any map $g:M\to \B\mathbb{G}$.

Clearly Claim 2 implies Claim 1 since if we can find abelian groups $K_{[0]}$ and $K_{[1]}$ such that 
\be
{\B K_{[0]} \ltimes \B^2 K_{[1]}}\to \B\mathbb{G}\stackrel{f}{\to} \B(\Spin\times_{\Z_2}(\SU(2)\times_{\Z_2}\Z_8))\times\B^2\Z_2 
\ee
is a fibration and $(fg)^*(a\cup\mathcal{P}_2(x_2))=0$ for any 5-manifold $M$ and any map $g:M\to \B\mathbb{G}$, then we can define $\mathbb{G}'$ which sits in a homotopy pullback square
\begin{widetext}
\be
\xymatrix{
{\B K_{[0]} \ltimes \B^2 K_{[1]}}\ar[r]&\B \mathbb{G}'\ar[r]^-{f'}\ar[d]&  \B \Spin\times \B\SU(2) \times\B \Z_8 \times\B^2\Z_2 \ar[d] \quad\quad\\
{\B K_{[0]} \ltimes \B^2 K_{[1]}}\ar[r]&\B \mathbb{G}\ar[r]^-f &\B (\Spin\times_{\Z_2}(\SU(2)\times_{\Z_2}\Z_8)) \times\B^2\Z_2.}
\ee
\end{widetext}

Then 
\be
{\B K_{[0]} \ltimes \B^2 K_{[1]}}\to \B\mathbb{G}'\stackrel{f'}{\to} \B\Spin\times\B\SU(2)\times\B\Z_8\times\B^2\Z_2
\ee
is a fibration and $(f'g')^*(\tilde{a}\cup\frac{\mathcal{P}_2(x_2)}{2})=0$ for any 5-manifold $M$ and any map $g':M\to \B\mathbb{G}'$.

Since $\H_i(\B\Spin,\Z)=0$ for $i=1,2,3$, $\H^2(\B\Spin,K_{[0]})=\H^3(\B\Spin,K_{[1]})=0$ by the universal coefficient theorem, similarly we have $\H^2(\B\SU(2),K_{[0]})=\H^3(\B\SU(2),K_{[1]})=0$, so in order to prove claim 2, we need only prove

Claim 3: We cannot find finite abelian groups $K_{[0]}$ and $K_{[1]}$ such that 
\bea
{\B K_{[0]} \ltimes \B^2 K_{[1]}}\to \B\mathbb{G}\stackrel{f}{\to} \B\Z_8\times\B^2\Z_2
\eea
is a fibration and $(fg)^*(\tilde{a}\cup\frac{\mathcal{P}_2(x_2)}{2})=0$ for any Spin 5-manifold $M$ and any map $g:M\to \B\mathbb{G}$.

We prove Claim 3 by finding a counterexample.

For $M=S^1\times S^2\times S^2$, let $a,b$ be the generators of $\H^2(S^2\times S^2,\Z_2)$, $c$ be the generator of $\H^1(S^1,\Z_8)$, let $h:M\to\B\Z_8\times\B^2\Z_2$ be given by $(c,a+b)$.
The lifting problem 
\bea
\xymatrix{&\B\mathbb{G}\ar[d]^f\\
S^1\times S^2\times S^2\ar[r]^-h\ar@{-->}[ru]^g&\B\Z_8\times\B^2\Z_2
}
\eea
has a solution, but $(c\mod2)\cup\frac{\mathcal{P}_2(a+b)}{2}\ne0$. 

In general, if $F\to E\stackrel{p}{\to} B\stackrel{q}{\to} \Sigma F$ is a fiber sequence, then $[M,F]\to[M,E]\stackrel{p_*}{\to} [M,B]\stackrel{q_*}{\to}[M,\Sigma F]$ is an exact sequence of abelian groups, so the lifting problem has a solution if and only if $q_*(h)=0$ where $q:\B\Z_8\times\B^2\Z_2\to
{\B^2 K_{[0]} \ltimes \B^3 K_{[1]}}$. So we need prove that $q\circ h=0$. 

{
Again apply the exact sequence $[M,\Sigma F]\to[M,\Sigma E]\to [M,\Sigma B]$ to $M=S^1\times S^2\times S^2$ and the fibration
\bea
\xymatrix{\B^2K_{[1]}\ar[r]&
{\B K_{[0]} \ltimes \B^2 K_{[1]}}\ar[d]\\
&\B K_{[0]},}
\eea
we get that if the image of $q\circ h$ in $[M,\Sigma B]$ is zero, then $q\circ h$ is the image of some map in $[M,\Sigma F]$.
}

{
We can write the composition $q'$ of the map $
{\B^2 K_{[0]} \ltimes \B^3 K_{[1]}}\to \B^2 K_{[0]}$ and $q$ as
\bea
q'=\left(\begin{array}{cc}q_1\\q_2\end{array}\right),
\eea
where $q_1\in\H^2(\B\Z_8,K_{[0]})$, $q_2\in\H^2(\B^2\Z_2,K_{[0]})$.
We assume that $q_2=0$ to ensure that the 1-form symmetry (here the 1-form $\Z_2^e$-symmetry) is not broken (write $f:\B\mathbb{G}\to\B\Z_8\times\B^2\Z_2$ as $f=(f_1,f_2)$, then 
$q_1\circ f_1+q_2\circ f_2=0$, if $q_2$ is nonzero, then $f_1$ and $f_2$ are locked, hence the 1-form symmetry is broken).
}

{We see that
$q'\circ h=q_1\circ c+q_2\circ(a+b)$, since $q_1\circ c=0$ and $q_2\circ (a+b)=0$ for $M=S^1\times S^2\times S^2$, so $q\circ h$ is the image of some map in $[M,\B^3K_{[1]}]$.
So $q\circ h=q_3\circ c+q_4\circ (a+b)$ where $q_3\in\H^3(\B\Z_8,K_{[1]})$, $q_4\in\H^3(\B^2\Z_2,K_{[1]})$. Since $q_3\circ c=0$ and $q_4\circ (a+b)=0$ for $M=S^1\times S^2\times S^2$, we have proven that $q\circ h=0$.
}

{In this appendix, we give a proof of the conclusion in the previous appendix. This answers the first question
(\ref{que1}) in Sec. \ref{sec:intro}.}

\section{Pullback trivialization of $\mathcal{P}_2(B_2)$ in $\Omega_4^{\SO}(\B^2\Z_2)$}

\label{sec:P(B)}

There is a group homomorphism:
\bea
\Omega_4^{\SO}(X)&\stackrel{\rho}{\to}&\Omega_4^{\SO}(\B^2\Z_2)\notag\\
(M,g)&\mapsto&(M,fg)
\eea

We want to extend the 1-form $\Z_2$ symmetry by 0-form $K_{[0]}$ symmetry and 1-form $K_{[1]}$ symmetry such that $\rho^*\mathcal{P}_2(\tilde{g})=\mathcal{P}_2(fg)=0$ for any $(M,g)\in\Omega_4^{\SO}(X)$ where $(M,\tilde{g})\in\Omega_4^{\SO}(\B^2\Z_2)$.

We consider the trivialization problem:
Does there exist a fibration $f:X\to \B^2\Z_2$ with fiber 
$
{\B K_{[0]} \ltimes \B^2 K_{[1]}}$ where $K_{[0]}$ and $K_{[1]}$ are finite abelian groups such that $\mathcal{P}_2(fg)=0$ for any oriented 4-manifold $M$ and any map $g:M\to X$?

The answer to this problem is negative, for $M=S^2\times S^2$, let $a,b$ be the generators of $\H^2(S^2\times S^2,\Z_2)$.
The lifting problem
\bea
\xymatrix{&X\ar[d]^f\\
S^2\times S^2\ar[r]^{a+b}\ar@{-->}[ru]^g&\B^2\Z_2
}
\eea
always has a solution, 
{but $\mathcal{P}_2(fg)=\mathcal{P}_2(a+b)$ is nontrivial.} Similarly as before, we need only prove that the composition map $S^2\times S^2\stackrel{a+b}{\to}\B^2\Z_2\stackrel{{q}}{\to}{\B^2 K_{[0]} \ltimes \B^3 K_{[1]}}$ is zero. 

{
This can be proven similarly as before.
}

{So $\mathcal{P}_2(x_2)$ cannot be trivialized.}

{In this appendix, we consider the pullback trivialization problem of $\mathcal{P}_2(x_2)$, we give a similar proof that $\mathcal{P}_2(x_2)$ also cannot be trivialized via extending the global symmetry by 0-form symmetry and 1-form symmetry.
This answers the second question
(\ref{que2}) in Sec. \ref{sec:intro}.}

\onecolumngrid
\bibliography{Yang-Mills-TQFT.bib, Yang-Mills-JW-2.bib} 

\end{document}